\documentclass{appolb}
\usepackage{epsfig}
\begin{document}
\title{Diffraction at HERA, the Tevatron and the LHC}
\author{C. Royon
\address{DAPNIA/Service de physique des particules, CEA/Saclay, 91191 
Gif-sur-Yvette cedex, France}}
\maketitle
\begin{abstract}
In these lectures, we present and discuss the most recent results on
inclusive diffraction from the HERA and Tevatron colliders and give the
prospects for the future at the LHC. Of special interest is the exclusive
production of Higgs boson and heavy objects ($W$, top, stop pairs) which will
require a better understanding of diffractive events and the link between $ep$
and hadronic colliders, as well as precise measurements and analyses of inclusive
diffraction at the LHC in particular to constrain further the gluon density in
the pomeron.
\end{abstract}

In these lectures, we describe the most recent results on inclusive
diffraction at HERA, as well as diffractive results from the Tevatron. We 
finish the lecture by discussing the prospects of diffractive physics at the LHC.

\section{Experimental definition of diffraction}

In this section, we discuss the different experimental ways to define
diffraction. As an example, we describe the methods used by the H1 and ZEUS
experiments at HERA, DESY, Hamburg in Germany. 

\subsection{The rapidity gap method}
HERA is a collider where electrons of 27.6 GeV collide with protons of 920 GeV.
A typical event as shown in the upper plot of Fig. 1 is $ep \rightarrow eX$
where electron and jets are produced in the final state. We
notice that the electron is scattered in the H1 backward detector\footnote{At
HERA, the backward (resp. forward) directions are defined as the direction
of the outgoing electron (resp. proton).} (in green)
whereas some hadronic activity is present in the forward region of the detector
(in the LAr calorimeter and in the forward muon detectors). The proton is thus
completely destroyed and the interaction leads to jets and proton remnants directly observable
in the detector. The fact that much energy is observed in the forward region is
due to colour exchange between the scattered jet and the proton remnants.
In about 10\% of the events, the situation is completely
different. Such events appear like the one shown in the bottom plot of Fig. 1.
The electron is still present in the backward detector, there is
still some hadronic activity (jets) in the LAr calorimeter, but no energy above
noise level is deposited in the forward part of the LAr calorimeter or in the
forward muon detectors. In other words, there is no color exchange between the
proton and the produced jets. As an example, this can be explained if the proton stays intact
after the interaction. 

This experimental observation leads to the first definition of diffraction:
request a
rapidity gap (in other words a domain in the forward detectors where  no
energy is deposited above noise level) in the forward region. For example, the H1
collaboration requests no energy deposition in the rapidity region
$3.3 < \eta < 7.5$ where $\eta$ is the pseudorapidity. Let us note that this
approach does not insure that the proton stays intact after the interaction, but
it represents a limit on the mass of the produced object $M_Y<1.6$ GeV. Within
this limit, the proton could be dissociated. The adavantage of the rapidity gap
method is that it is quite easy to implement and it has a large acceptance in 
the diffractive kinematical
plane.

\begin{figure}[t]
\begin{center}
\vspace{10.cm}
\hspace{-4cm}
\epsfig{file=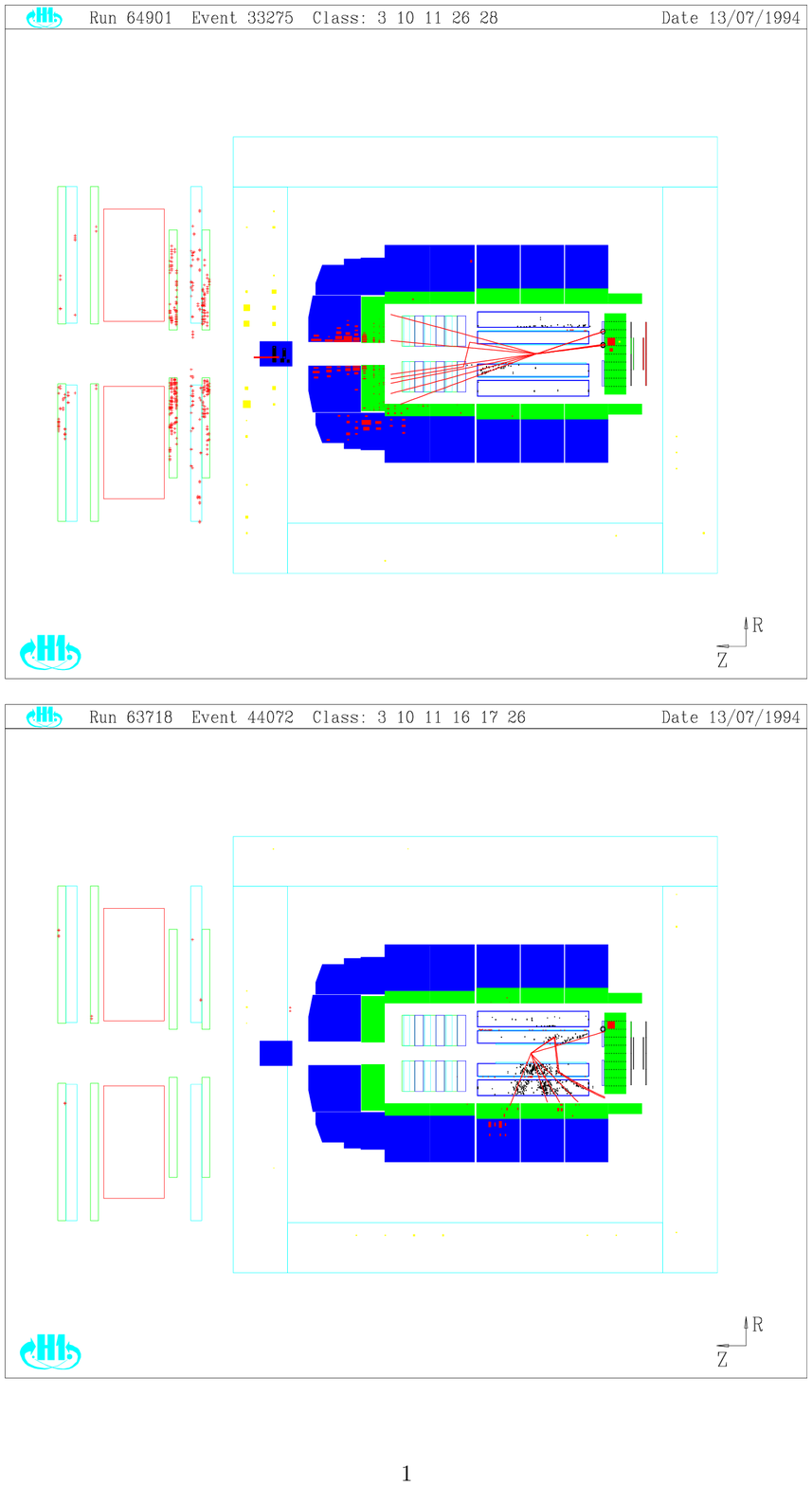,width=5.2cm}
\caption{``Usual" and diffractive events in the H1 experiment.}
\end{center}
\label{fig1}
\end{figure}

\subsection{Proton tagging}
The second experimental method to detect diffractive events is also natural: the
idea is to detect directly the intact proton in the final state. The proton
loses a small fraction of its energy and is thus scattered at very small angle
with respect to the beam direction. Some special detectors called roman pots can
be used to detect the protons close to the beam. The basic idea is simple: the roman pot
detectors are located far away from the interaction point and can move close to
the beam, when the beam is stable, to detect protons scattered at vary small
angles. The inconvenience is that the kinematical reach of those detectors is
much smaller than with the rapidity gap method. On the other hand,
the advantage is that it
gives a clear signal of diffraction since it measures the diffracted proton
directly.

A scheme of a roman pot detector as it is used by the H1 or ZEUS experiment is shown
in Fig. 2. The beam is the horizontal line at the upper part of the
figure. The detector is located in the pot itself and can move closer to the
beam when the beam is stable enough (during the injection period, the detectors
are protected in the home position). Step motors allow to move the detectors
with high precision. A precise knowledge  of the detector position is
necessary to reconstruct the transverse momentum of the scattered proton and
thus the diffractive kinematical variables. The detectors are placed in a
secondary vaccuum with respect to the beam one. 

\begin{figure}[t]
\begin{center}
\epsfig{file=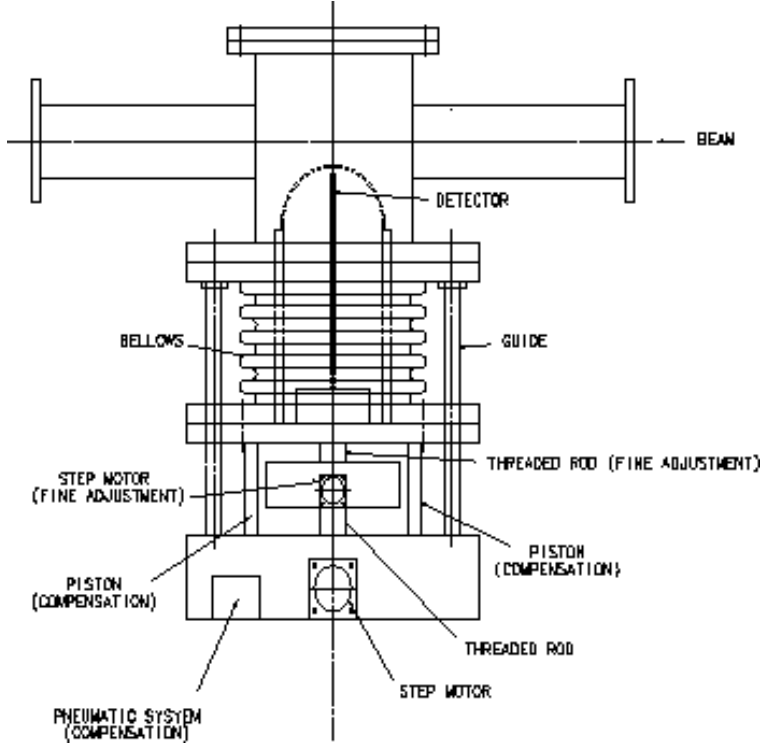,width=10cm}
\caption{Scheme of a roman pot detector.}
\end{center}
\label{fig2}
\end{figure} 

\subsection{The $M_X$ method}
The third method used at HERA mainly by the ZEUS experiment is slightly less
natural. It is based on the fact that there is a different behaviour in $\log
M_X^2$ where $M_X$ is the total invariant mass produced in the event either 
for diffractive or
non diffractive events. For diffractive events $d \sigma_{diff}/dM_X^2 =  
(s/M_X^2)  ^{\alpha -1} = const. ~~$if$ ~~ \alpha \sim 1$ (which is the case for
diffractive events). The ZEUS collaboration performs some fits of the 
$d\sigma/dM_X^2$ distribution:
\begin{eqnarray}
\frac{d \sigma}{dM_X^2} = D + c \exp(b \log M_X^2)
\end{eqnarray}
as illustrated in Fig. 3. The usual non diffractive events are
exponentially suppressed at high values of $M_X$. The difference between the
observed $d\sigma/dM_X^2$ data and the exponential suppressed distribution is
the diffractive event contribution. This method, although easy to implement, presents
the inconvenience that it relies strongly on the assumption of
the exponential suppression of non diffractive events.

\begin{figure}[t]
\begin{center}
\hspace{11cm}
\epsfig{file=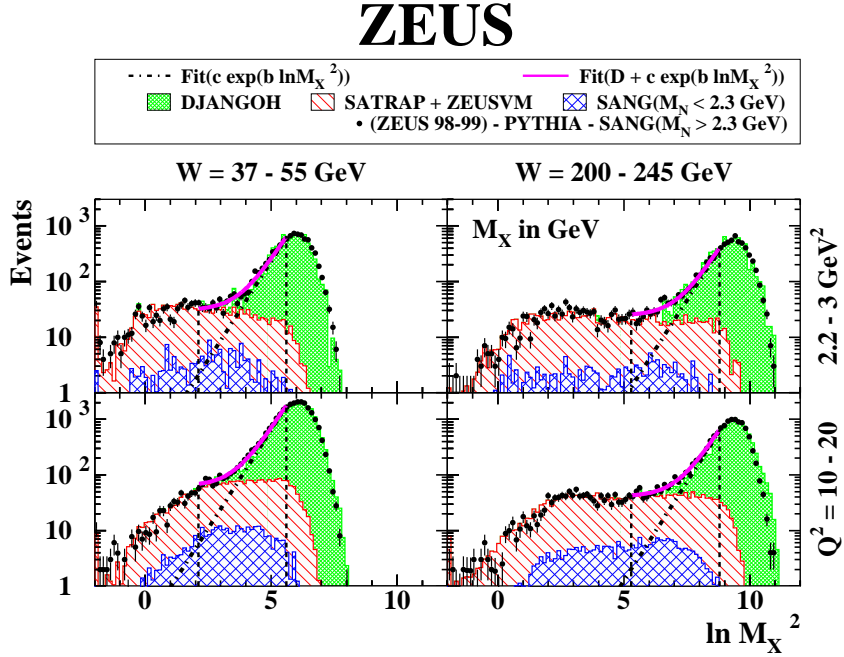,width=10cm, angle=270}
\caption{Illustration of the $M_X$ method used by the ZEUS collaboration to
define diffractive events.}
\end{center}
\label{fig3}
\end{figure}

\subsection{Diffractive kinematical variables}
After having described the different experimental definitions of diffraction at
HERA, we will give the new kinematical variables used to characterise diffraction.
A typical diffractive event is shown in Fig. 4 where $ep \rightarrow
epX$ is depicted. In addition to the usual deep inelastic variables, $Q^2$ the transfered
energy squared at the electron vertex, $x$ the fraction of the proton momentum
carried by the struck quark, 
$W^2 = Q^2 (1/x -1)$ the total energy in the final state,
new diffractive variables are defined: $x_P$ is the
momentum fraction of the proton carried by the colourless object called the
pomeron, and $\beta$ the momentum fraction of the pomeron carried by the
interacting parton inside the pomeron if we assume the pomeron to be made of
quarks and gluons:
\begin{eqnarray}
x_P &=& \xi = \frac{Q^2+M_X^2}{Q^2+W^2} \\
\beta &=& \frac{Q^2}{Q^2+M_X^2} = \frac{x}{x_P}.
\end{eqnarray}

\begin{figure}[t]
\begin{center}
\epsfig{file=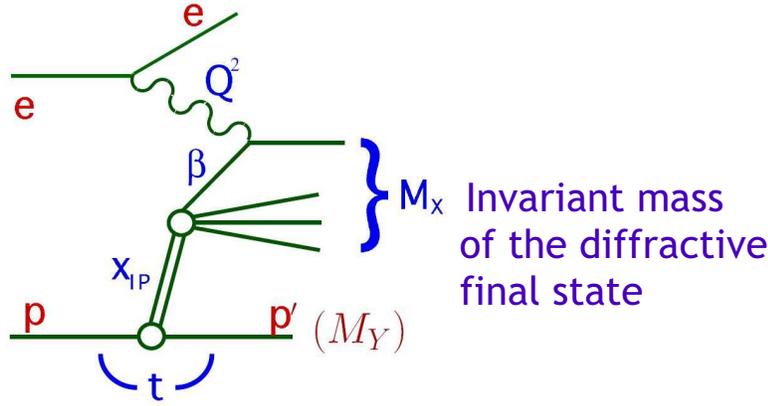,width=6cm,angle=270}
\caption{Scheme of a diffractive event at HERA.}
\end{center}
\label{fig4}
\end{figure}

\section{Diffractive structure function measurement at HERA}

\subsection{Diffractive factorisation}
In the following diffractive structure function analysis, we 
distinguish two kinds of factorisation at HERA. The first factorisation is the
QCD hard scattering collinear factorisation at fixed $x_P$ and $t$
(see left plot of Fig. 5)
\cite{collins}, namely

\begin{eqnarray}
d \sigma (ep \rightarrow eXY) = f_D(x,Q^2,x_P,t) \times
d \hat{\sigma} (x,Q^2)
\end{eqnarray}
where we can factorise the flux $f_D$ from the cross section $\hat{\sigma}$.
This factorisation was proven recently, and separates the $\gamma q$ coupling to
the interaction with the colourless object. 

The Regge factorisation at the proton vertex allows to factorise 
the $(x_P,t)$ and $(\beta,Q^2)$ dependence, or in other words the hard
interaction from the pomeron coupling to the proton (see right plot of Fig. 5).

\begin{figure}[t]
\begin{center}
\epsfig{file=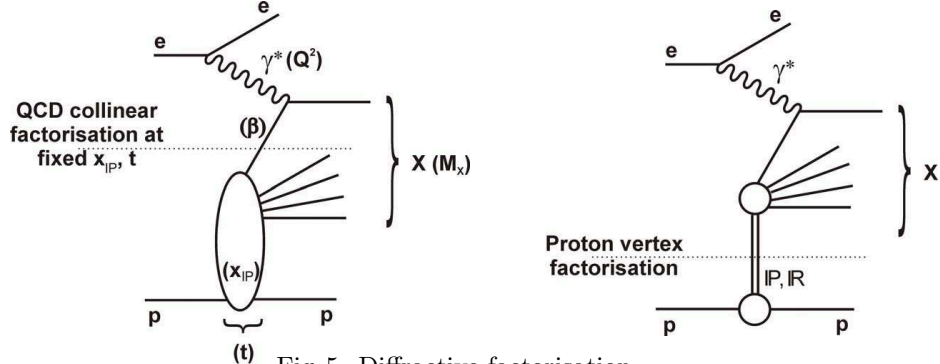,width=5cm,angle=270}
\caption{Diffractive factorisation}
\end{center}
\label{fig5}
\end{figure} 

\subsection{Measurement of the diffractive proton structure function}
The different measurements are performed using the three different methods to
define diffractive events described in the first section. As an example, the H1
collaboration measures the diffractive cross section $\sigma^D$ using the
rapidity gap method:

\begin{eqnarray}
\frac{d^3 \sigma^D}{d x_P dQ^2 d \beta} = \frac{2 \pi \alpha_{em}^2}{\beta Q^4}
\left( 1-y+\frac{y^2}{2} \right) \sigma_r^D(x_P, Q^2, \beta)
\end{eqnarray}
where $\sigma_r^D$ is the reduced diffractive cross section. 
The measurement \cite{h1f2d} is presented in Fig. 6.
We notice that the measurement has been performed with high precision over a
wide kinematical domain: $0.01 < \beta < 0.9$, $3.5 < Q^2 < 1600$ GeV$^2$,
$10^{-4}<x_P<5.10^{-2}$. The data are compared to the result of a QCD fit which
we will discuss in the following.

The rapidity gap data are also compared with the data obtained either using the
$M_X$ method or the one using proton tagging in roman pot detectors.
Since they do not correspond exactly to the same definition of diffraction, a
correction factor of 0.85 must be applied to the ZEUS $M_X$ method to be
compared to the rapidity gap one (this factor is due to the fact that the two
methods correspond to two different regions in $M_Y$, namely $M_Y<1.6$ GeV for
H1 and $M_Y<2.3$ GeV for ZEUS). It is also possible to measure directly in the H1
experiment the ratio of the diffractive structure function measurements between
the rapidity gap and the proton tagging methods as illustrated in Fig.
7. Unfortunately, the measurement using the proton tagging method is
performed only in a restricted kinematical domain. No
kinematical dependence has been found within uncertainties for this ratio
inside this kinematical domain (see Fig. 7 for
the $\beta$ and $Q^2$ dependence, and Ref. \cite{f2dh1tag} for the $x_P$
dependence as well). Note that the ratio could still be depending on
$\beta$ and $Q^2$ outside the limited domain of  measurement.

\newpage


\begin{figure}[t]
\begin{center}
\epsfig{file=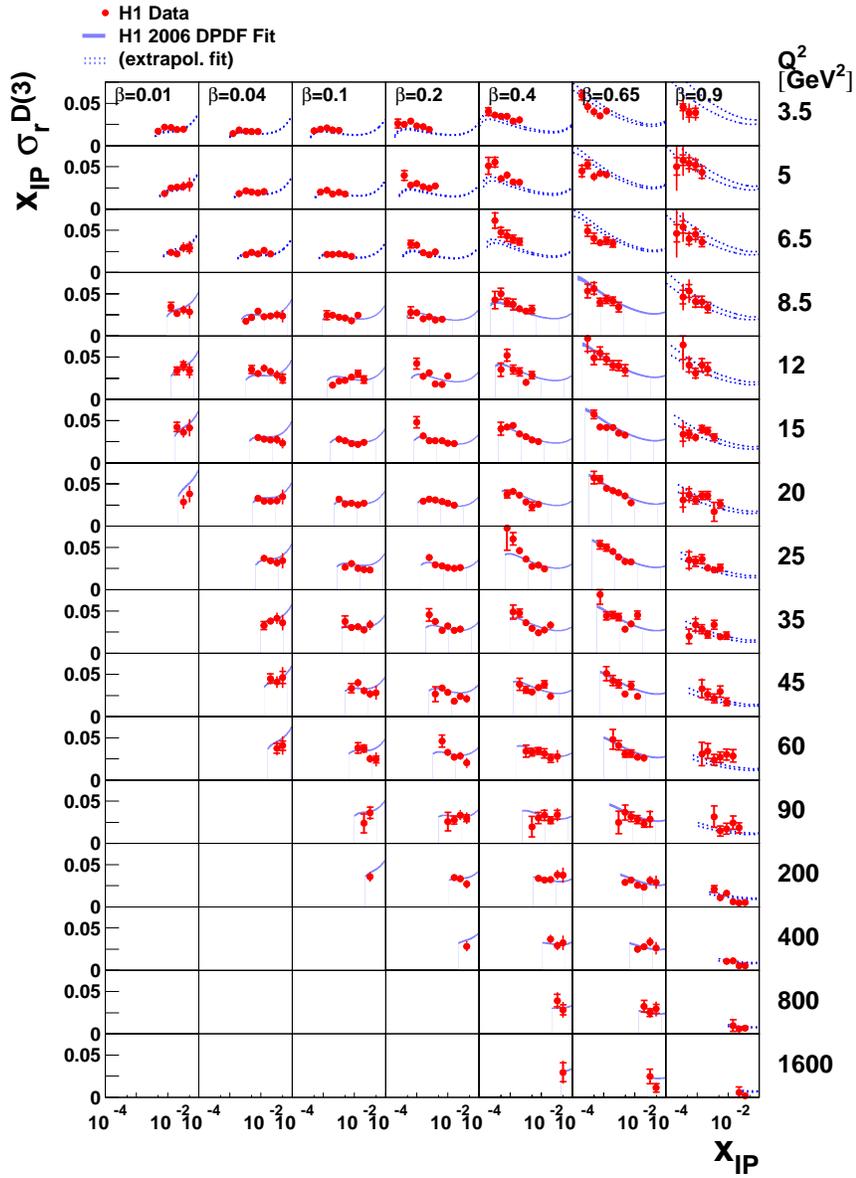,width=12cm}
\vspace{17cm}
\caption{Measurement of the diffractive structure function by the H1
collaboration}
\end{center}
\label{fig6}
\end{figure} 

\newpage

\begin{figure}[t]
\begin{center}
\begin{tabular}{cc}
\hspace{-1cm}
\epsfig{figure=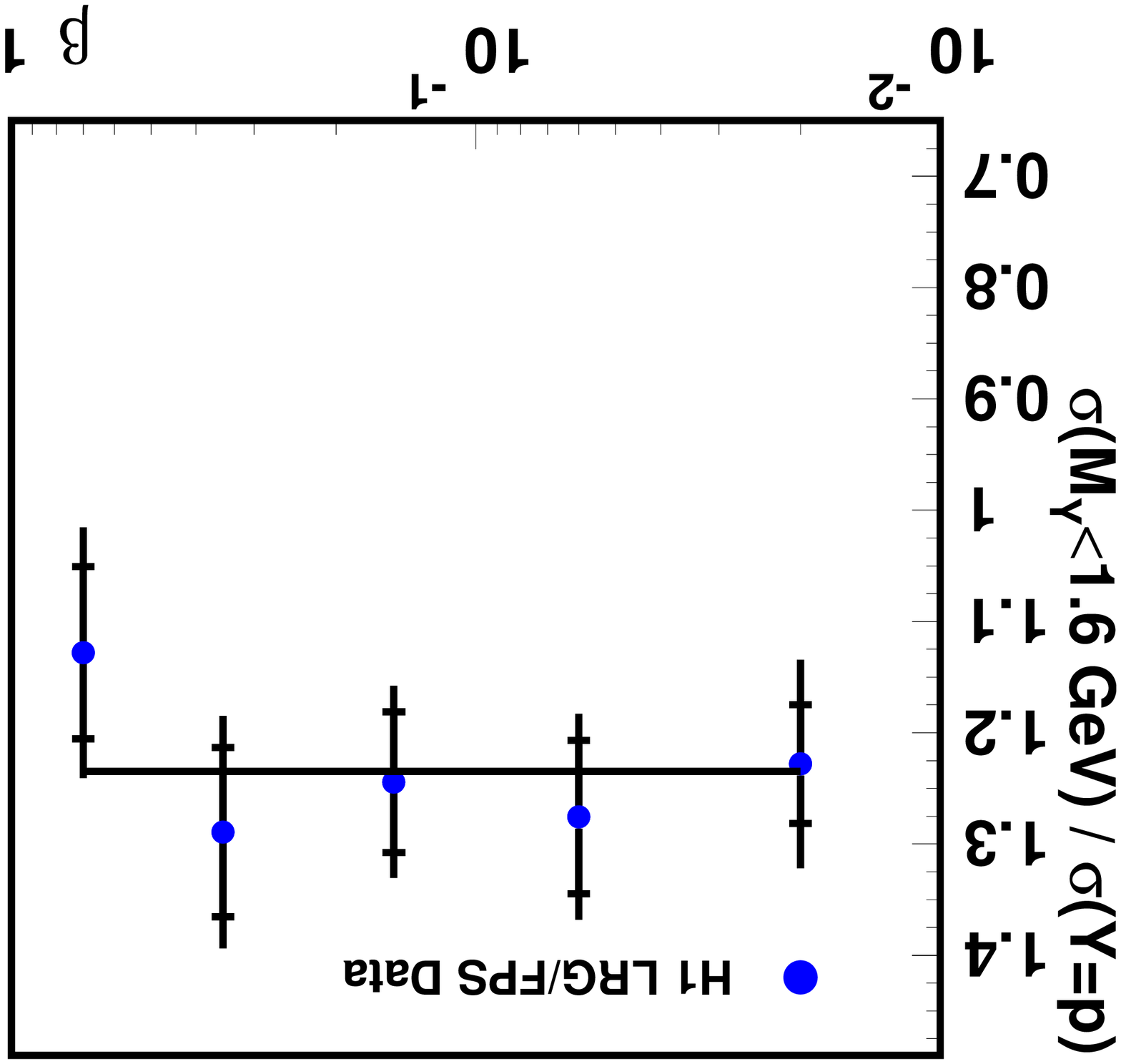,height=2.0in,angle=180} &
\hspace{11cm}
\epsfig{figure=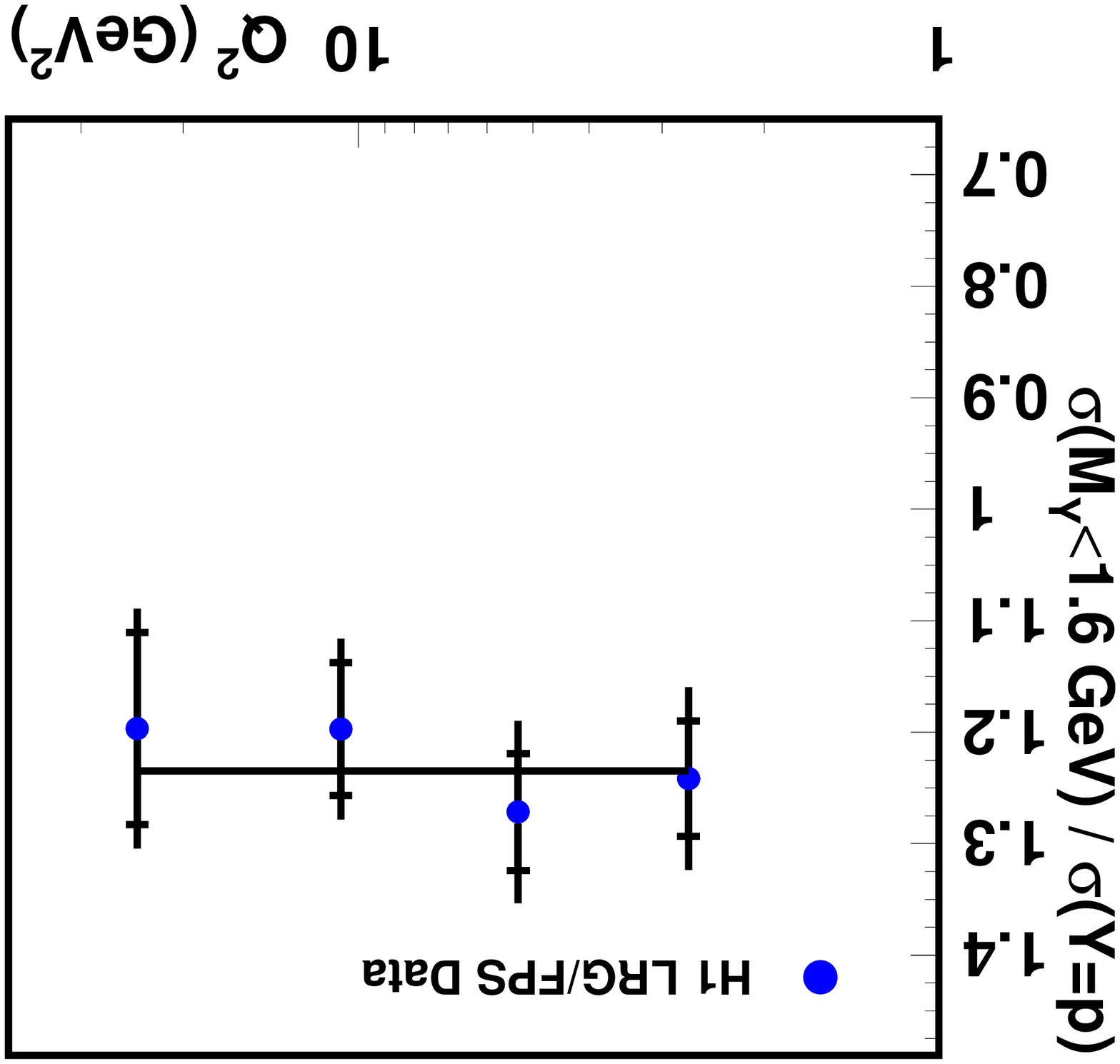,height=2.0in,angle=180} \\
\end{tabular}
\caption{Measurement of the ratio of the diffractive structure function
between the rapidity gap and the proton tagging methods (H1 experiment).}
\end{center}
\label{fig7}
\end{figure}

\subsection{QCD analysis of the diffractive structure function measurement}
As we mentionned already, according to Regge theory, we can factorise the 
$(x_P,t)$ dependence from the $(\beta,Q^2)$ one. The first  diffractive structure
function measurement from the H1 collaboartion \cite{firstf2d} showed that this
assumption was not true. The natural solution as observed in soft
physics was that two different
trajectories, namely pomeron and secondary reggeon, were needed to describe the
measurement, which lead to a good description of the data. The diffractive structure
function then reads:

\begin{eqnarray}
F_2^D \sim f_p(x_P) (F_2^D)_{Pom}(\beta, Q^2) + 
f_r(x_P) (F_2^D)_{Reg}(\beta, Q^2)
\end{eqnarray}
where $f_p$ and $f_r$ are the pomeron and reggeon fluxes, and $(F_2^D)_{Pom}$
and $(F_2^D)_{Reg}$ the pomeron and reggeon structure functions. The flux
parametrisation is predicted by Regge theory:

\begin{eqnarray}
f(x_P,t) = \frac{e^{B_Pt}}{x_P^{2 \alpha_P(t) -1}}
\end{eqnarray}
with the following pomeron trajectory

\begin{eqnarray}
\alpha_P (t)= \alpha_P(0) + \alpha'_P t.
\end{eqnarray}
The $t$ dependence has been obtained using the proton tagging method, and the
following values have been found: $\alpha'_P = 0.06^{+0.19}_{-0.06}$ GeV$^{-2}$,
$B_P=5.5^{+0.7}_{-2.0}$ GeV$^{-2}$ (H1). Similarly, the values of $\alpha_P(0)$
have been measured using either the rapidity gap for H1 or the $M_X$ method for
ZEUS in the QCD fit described in the next paragraph \cite{us,h1f2d}. 
The Reggeon parameters have been found to be $\alpha'_R = 0.3$ GeV$^{-2}$,
$B_R=1.6$ GeV$^{-2}$ (H1). The value of $\alpha_R(0)$ has been determined from
rapidity gap data and found to be equal to 0.5. Since the reggeon is expected
to have a similar $q \bar{q}$ structure as the pion and the data are poorly
sensitive to the structure function of the secondary reggeon, it was assumed to
be similar to the pion structure with a free normalisation.

The next step is to perform Dokshitzer Gribov Lipatov Altarelli Parisi (DGLAP)
\cite{dglap} fits to the pomeron structure function. If we assume that the
pomeron is made of quarks and gluons, it is natural to check whether the DGLAP
evolution equations are able to describe the $Q^2$ evolution of these parton
densities. As necessary for DGLAP fits, a form for the input distributions is assumed
at a given $Q_0^2$ and is evolved using the DGLAP evolution equations to a
different $Q^2$, and fitted to the diffractive structure function data at
this $Q^2$ value. The form of the distribution at $Q_0^2$ has been chosen to be:

\begin{eqnarray}
\beta q &=& A_q \beta^{B_q} (1-\beta)^{C_q}  \\
\beta G &=& A_g  (1-\beta)^{C_g},
\end{eqnarray}
leading to three (resp. two) parameters for the quark (resp. gluon) densities.
At low $\beta$, the evolution is driven by $g \rightarrow q \bar{q}$ while
$q  \rightarrow qg$ becomes more important at high $\beta$. All diffractive data
with $Q^2 > 8.5$ GeV$^2$ and $\beta<0.8$ have been used in the fit 
\cite{h1f2d,us} (the high
$\beta$ points being excluded to avoid the low mass region where the vector
meson resonances appear). This leads to a good description of all diffractive
data included in the fit.

The DGLAP QCD fit allows to get the parton distributions in the pomeron as a
direct output of the fit, and is displayed in Fig. 8 as a blue shaded
area as a function of $\beta$. We first note that the gluon density is much
higher than the quark one, showing that the pomeron is gluon dominated. We also
note that the gluon density at high $\beta$ is poorly constrained which is shown
by the larger shaded area. 

Another fit was also performed by the H1 collaboration imposing $C_g=0$.
While the fit quality is similar, the gluon at high $\beta$ is quite different,
and is displayed as a black line in Fig. 8 ($z$ is the equivalent of $\beta$ for
quarks). This shows further that the
gluon is very poorly constrained at high $\beta$ and some other data sets such
as jet cross section measurements are needed to constrain it further.

\begin{figure}[t]
\begin{center}
\epsfig{file=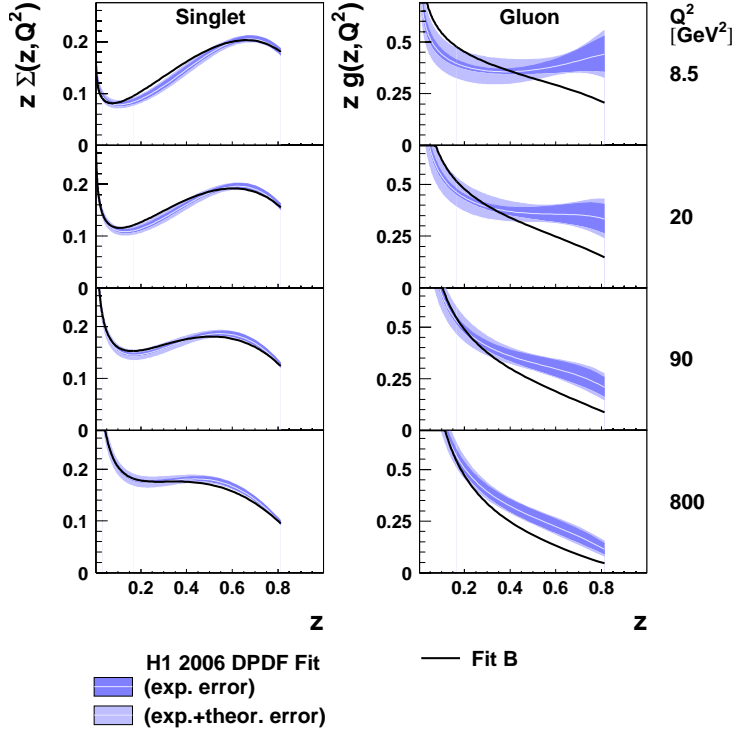,width=10cm}
\vspace{12cm}
\caption{Extraction of the parton densities in the pomeron using a DGLAP NLO fit
(H1 collaboration).}
\end{center}
\label{fig8}
\end{figure} 

\subsection{QCD fits using diffractive structure function and jet cross
section measurements}

In this section, we describe combined fits using diffractive structure function
and jet cross section data to further constrain the gluon at high $\beta$.
First, it is possible to compare the diffractive dijet cross section measurements
with the predictions using the gluon and quark densities from the QCD fits
described in the previous section. The comparison \cite{h1f2d} shows a 
discrepancy between the measurement and the expectation from the QCD fit by
about a factor 2 at high $\beta$. This motivates the fact that it is important
to add the jet cross section data to the inclusive structure function
measurement in the QCD fit to further constrain the gluon density at high
$\beta$. The new parton distributions are shown in Fig. 9 as a blue
shaded area. The comparison between the jet cross section measurements and the
prediction from the QCD fits are in good agreement as shown in Fig. 10.
The present uncertainty is of the order of 50\% at high $\beta$.

\begin{figure}[t]
\begin{center}
\epsfig{file=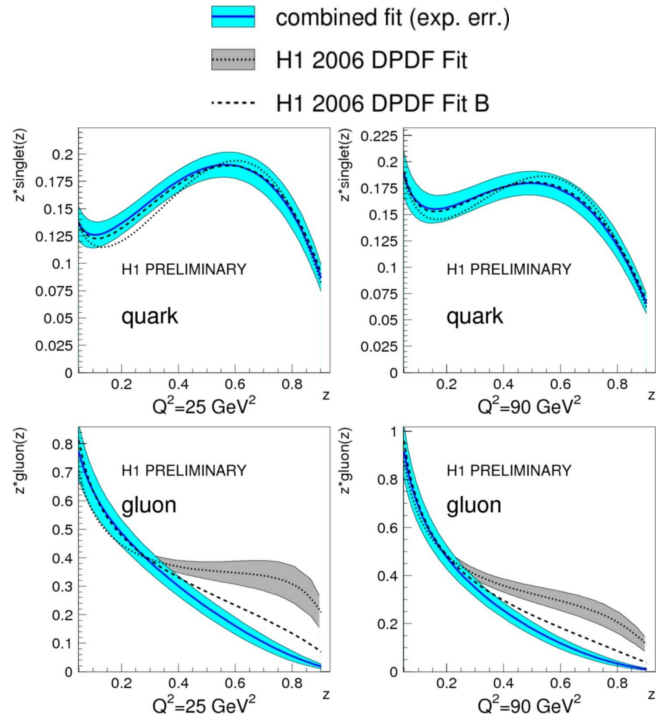,width=10cm}
\vspace{12cm}
\caption{Extraction of the parton densities in the pomeron using a DGLAP NLO fit
(H1 collaboration). The blue shaded area shows the results after including both
the diffractive structure function and the dijet cross section measurements into
the QCD DGLAP fit.}
\end{center}
\label{fig9}
\end{figure}

\begin{figure}[t]
\begin{center}
\epsfig{file=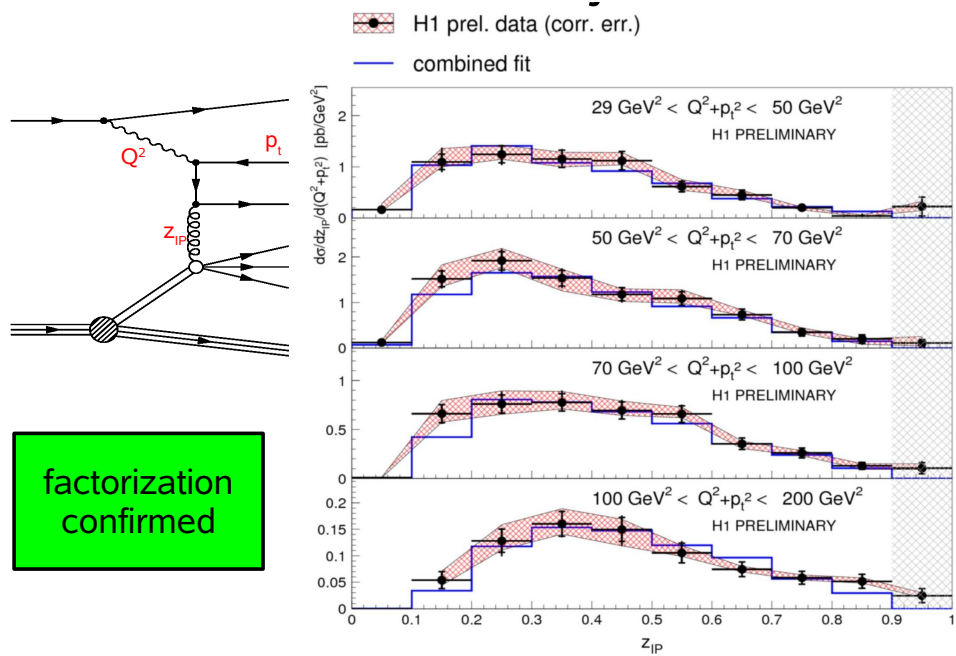,width=9cm,angle=270}
\caption{Comparison between the H1 QCD diffractive fit based on diffractive
structure function and dijet data and the dijet data.}
\end{center}
\label{fig10}
\end{figure} 

\subsection{Other models describing inclusive diffraction at HERA}
Many different kinds of models can be used to describe inclusive diffraction at
HERA, and we will describe here only
the results based on the two gluon model \cite{bekw}. Other models of interest
such as the BFKL dipole model \cite{dipole} or the saturation model 
\cite{saturation} are described in Ref. \cite{us} as well as the results of the
fits to the diffractive data. Due to the lack of time, we cannot describe them
in these lectures. 

The 2-gluon model \cite{bekw} starts from the image of a perturbative pomeron
made of two gluons and coupled non perturbatively to the proton. 
As shown in Fig. 11,
there are three main contributions to the diffractive structure function, namely
the $q \bar{q}$ transverse, $q \bar{q} g$ (neglecting higher order Fock states)
and the $q \bar{q}$ longitudinal terms. Contrary to the QCD fits described in the previous
section, there is no concept of  diffractive PDFs in this approach.
The $\beta$-dependence of the structure function is motivated by some general features 
of QCD-parton model calculations: at small $\beta$ 
the spin 1/2 (quark) exchange in the $q\bar{q}$ production leads to a
behaviour $\sim \beta$, whereas the spin 1 (gluon) exchange in the
$q \bar{q} g$ term corresponds 
to $\beta^0$. For large $\beta$, perturbative QCD leads to 
$1-\beta$ and $(1-\beta)^0$ for the transverse and longitudinal
$q \bar{q} $ terms respectively. 
Concerning the $Q^2$ dependence, the longitudinal term is a higher twist
one. Finally, the dependence on $x_P$ cannot be obtained
from perturbative QCD and therefore is left free. 
An additional sub-leading trajectory (secondary reggeon) has to be
parametrised from soft physics and
is added to the model
as for the DGLAP based fit to describe H1 data.

The 2-gluon model leads to a good description of both ZEUS and H1 data. As an
example, the comparison of the ZEUS $M_X$ 
data \cite{zeusf2d} in different $x_P$ and $Q^2$ bins as a function  of $\beta$  
with the 2-gluon model is given
in Fig. 12 where we note the good agreement between the model and the
data. Fig. 12 also describes independently the three components of the
model, namely the tranverse $q \bar{q}$ one which dominates at medium $\beta$,
the $q \bar{q} g$ one at low $\beta$, and the longitudinal higher twist $q
\bar{q}$ one at high $\beta$. 

\begin{figure}[t]
\begin{center}
\epsfig{file=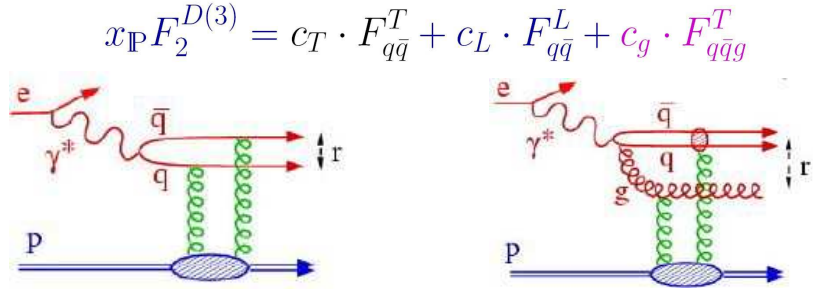,width=5cm,angle=270}
\caption{Schematic view of the 2 gluon model \cite{bekw}.}
\end{center}
\label{fig11}
\end{figure}

\begin{figure}[t]
\begin{center}
\vspace{-3cm}
\hspace{-4cm}
\epsfig{file=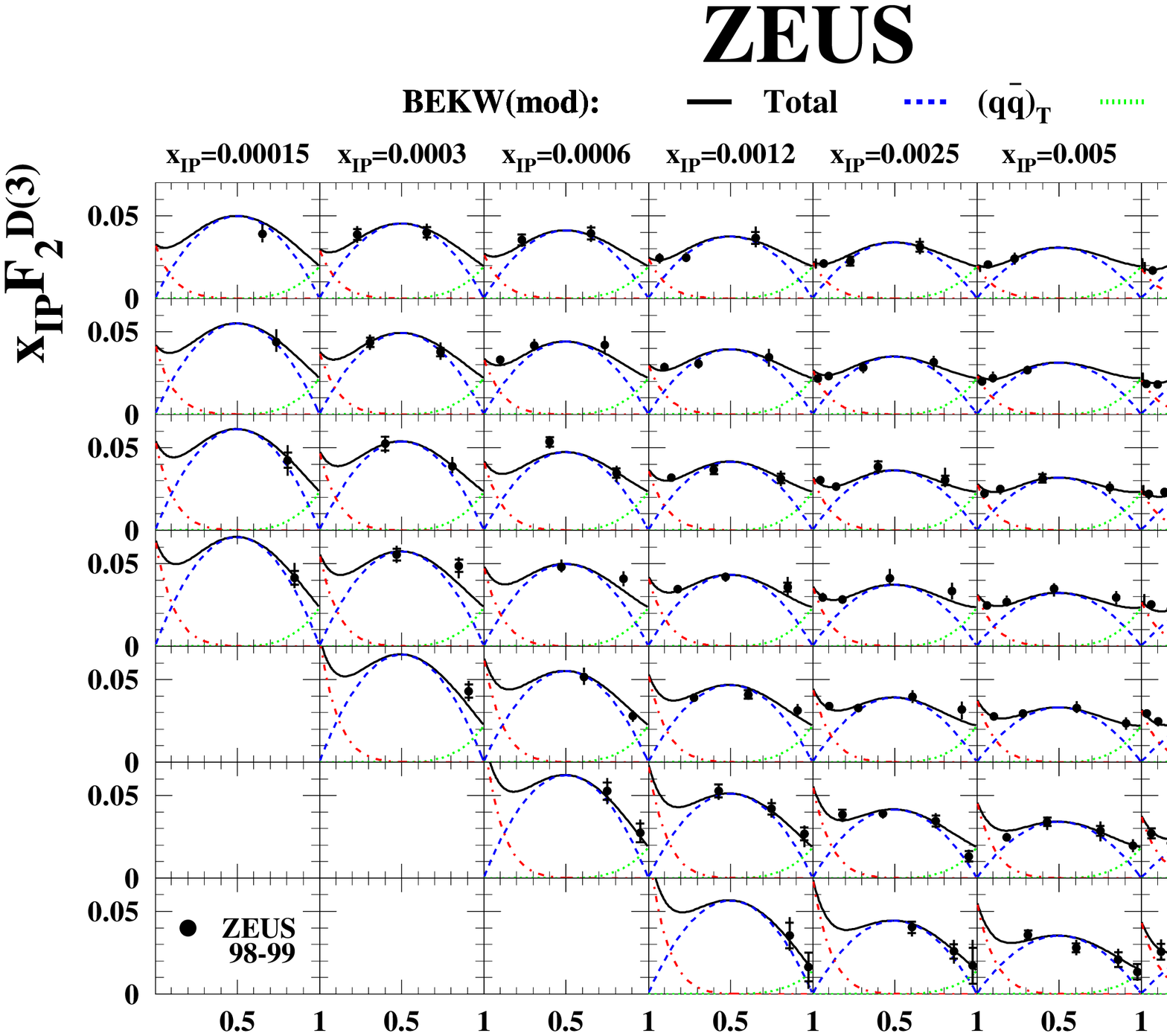,width=10cm}
\caption{Comparison between the 2 gluon model with the ZEUS $M_X$ data.}
\end{center}
\label{fig12}
\end{figure}

\section{Diffraction at the Tevatron}

The Tevatron is a $p \bar{p}$ collider located close to Chicago at Fermilab,
USA. It is presently the collider with the highest center-of-mass energy of
about 2 TeV. Two main experiments are located around the ring, D\O\ and CDF. 
Both collaborations have accumulated a luminosity of the order of 1.5 fb$^{-1}$ with an efficiency
of about 85\%.

\subsection{Diffractive kinematical variables}

The difference between diffraction at HERA and at the Tevatron is that
diffraction can occur not only on either $p$ or $\bar{p}$ side as at 
HERA, but also on both sides. The former case is called single diffraction
whereas the other one double pomeron exchange. In the same way
as we defined the kinematical variables $x_P$ and $\beta$ at HERA, we define $\xi_{1,2}$(=$x_P$ at HERA) 
as the proton fractional momentum loss (or as the $p$ or
$\bar{p}$ momentum fraction carried by the pomeron), and $\beta_{1,2}$, the fraction of the
pomeron momentum carried by the interacting parton. The produced diffractive
mass is equal to $M^2= s \xi_1 $ for single diffractive events and to
$M^2= s \xi_1 \xi_2$ for double pomeron exchange. The size of the rapidity gap
is of the order of $\Delta \eta \sim \log 1/ \xi_{1,2}$.

\subsection{How to find diffractive events at the Tevatron?}
The selection of diffractive events at the Tevatron follows naturally from the
diffractive event selection at HERA. The D\O\ and CDF collaborations obtained
their first diffractive results using the rapidity gap method which showed that
the percentage of single diffractive events was of the order of 1\%, and about
0.1\% for double pomeron exchanges. Unfortunately, the reconstruction of the
kinematical variables is less precise than at HERA if one uses the rapidity gap
selection since it suffers from the worse resolution of reconstructing hadronic
final states.

The other more precise method is to tag directly the $p$ and $\bar{p}$ in the
final state. The CDF collaboration installed roman pot detectors in the outgoing
$\bar{p}$ direction only at the end of Run I \cite{cdfpots}, whereas the D\O\ collaboration
installed them both in the outgoing $p$ and $\bar{p}$ directions \cite{d0pots}. 
The D\O\ (dipole detectors) and CDF roman
pots cover the acceptance of $t$ close to 0 and $0.02 < \xi < 0.05$ in the
outgoing $\bar{p}$ direction only. In addition, the D\O\ coverage extends for
$0.5 < |t| < 1.5$ GeV$^2$, and $0.001< \xi < 0.03$ in both $p$ and $\bar{p}$
directions (quadrupole detectors).
The CDF collaboration completed the detectors in the forward region by adding a
miniplug calorimeter on both $p$ and $\bar{p}$ sides allowing a coverage of
$3.5 < |\eta| < 5.1$ and some beam showing counters close to beam pipe ($5.5 <
|\eta| < 7.5$) allowing to reject non diffractive events.

\subsection{Measurement of elastic events at D\O\ }
Due to the high value of the production cross section, one of the first physics
topics studied by the D\O\ collaboration was the elastic scattering cross
section. Elastic events can also be used to align precisely the detectors.
During its commissioning runs, the D\O\ collaboration was able to measure the
diffractive slope for elastic events using double tagged events. The D\O\
results together with the results from the previous lower energy experiments are
diplayed in Fig. 13. The normalisation of the D\O\ data is arbitrary
since the data were taken using the commissioning runs of the roman pot
detectors in stand-alone mode without any access to luminosity measurements.
These data show the potential of the D\O\ roman pot detectors and this
measurement will be performed again soon now that the roman pot detectors are
fully included in the D\O\ readout system. A great challenge is to measure the
change of slope in $t$ of the elastic cross section towards 0.55-0.6 GeV$^2$ 
predicted by the models. Many measurements such as the pomeron structure in single
diffractive events or double pomeron exchange, inclusive diffraction,
diffractive $Z$, $W$ and $b$-jets are being pursued in the D\O\ collaboration.

\begin{figure}[t]
\begin{center}
\epsfig{file=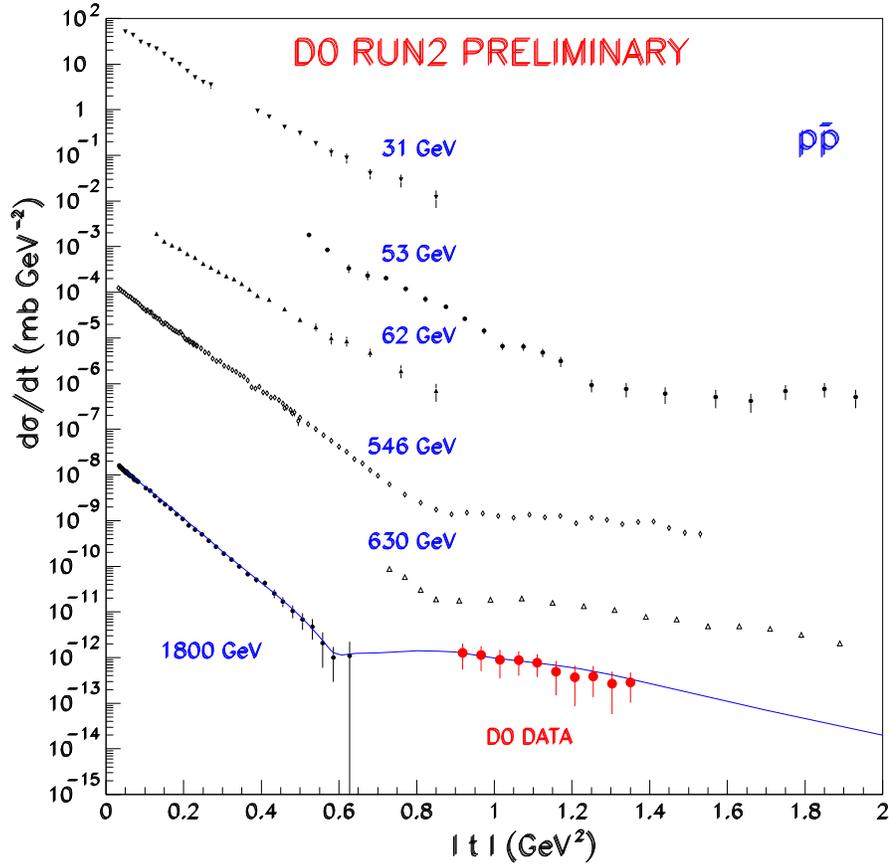,width=12cm}
\caption{Measurement of the $t$-slope of the elastic cross section in D\O\ .}
\end{center}
\label{fig13}
\end{figure}

\subsection{Factorisation or factorisation breaking at the Tevatron?}
The CDF collaboration measured diffractive events at the Tevatron and their
characteristics. In general, diffractive events show as expected less QCD
radiation: for instance, dijet events are more back-to-back or the difference in
azimuthal angles between both jets is more peaked towards $\pi$. To make
predictions at the Tevatron and the LHC, it is useful to know if factorisation
holds. In other words, is it possible to use the parton distributions in the pomeron obtained in
the previous section using HERA data to make predictions at the Tevatron, and
also further constrain the parton distribution functions in the pomeron since
the reach in the diffractive kinematical plane at the Tevatron and HERA is
different? Theoretically, factorisation is not expected to hold between the
Tevatron and HERA due to additional $pp$ or $p \bar{p}$ interactions. 
For instance, some soft gluon
exchanges between protons can occur at a longer time scale than the hard
interaction and destroy the rapidity gap or the proton does not remain intact
after interaction. The factorisation break-up is confirmed by comparing the percentage of
diffractive events at HERA and the Tevatron (10\% at HERA and about 1\% of
single diffractive events at the Tevatron) showing already that factorisation
does not hold. This introduces the concept of gap survival probability, the
probability that there is no soft additional interaction or in other words that
the event remains diffractive. We will mention in the following how this concept
can be tested directly at the Tevatron.

The first factorisation test  concerns CDF data
only. It is interesting to check whether factorisation holds within CDF data
alone, or in other words if the $\beta$ and $Q^2$ dependence can be factorised
out
from the $\xi$ one. Fig. 14 shows the percentage of diffractive events
as a function of $x$ for different $\xi$ bins and shows the same $x$-dependence
in all $\xi$ bins supporting the fact that CDF data are consistent with factorisation
\cite{cdfdiff}. The CDF collaboration also studied the $x$ dependence for
different $Q^2$ bins which lead to the same conclusions. This also shows that
the Tevatron data do not require additional secondary reggeon trajectories as in
H1.
 
The second step is to check whether factorisation holds or not between Tevatron and
HERA data. The measurement of the diffractive structure function is possible
directly at the Tevatron. The CDF collaboration measured the ratio of dijet
events in single diffractive and non diffractive events, which is directly
proportional to the ratio of the diffractive to the ``standard" proton structure
functions $F_2$:
\begin{eqnarray}
R(x) = \frac{Rate^{SD}_{jj} (x)}{Rate^{ND}_{jj} (x)} \sim
\frac{F^{SD}_{jj} (x)}{F^{ND}_{jj} (x)}
\end{eqnarray}
The ``standard" proton structure function is known from the usual PDFs obtained
by the
CTEQ or MRST collaborations. The comparison between the CDF measurement 
(black points, with systematics errors as shaded area) and the
expectation from the H1 QCD fits in full line is shown in Fig. 15. 
We notice a discrepancy of a factor 8 to 10 between the data and the predictions from
the QCD fit, showing that factorisation does not hold. However, the difference
is compatible with a constant on a large part of the kinematical plane in
$\beta$, showing that the survival probability does not seem to be
$\beta$-dependent within experimental uncertainties.

The other interesting measurement which can be also performed at the Tevatron is
the test of factorisation between single diffraction and double pomeron
exchange. The results from the CDF collaboration are shown in Fig. 16.
The left plot shows the definition of the two ratios while the right figure
shows the comparison between the ratio of double pomeron exchange to single
diffraction and the QCD predictions using HERA data in full line. Whereas
factorisation was not true for the ratio of single diffraction to non diffractive events,
factorisation holds for the ratio of double pomeron exchange to single
diffraction! In other words, the price to pay for one gap is the same as the
price to pay for two gaps. The survival probability, i.e. the probability not to emit an additional
soft gluon after the hard interaction needs to be applied only once to require
the existence of a diffractive event, but should not be applied again for double
pomeron exchange.

\begin{figure}[t]
\begin{center}
\epsfig{file=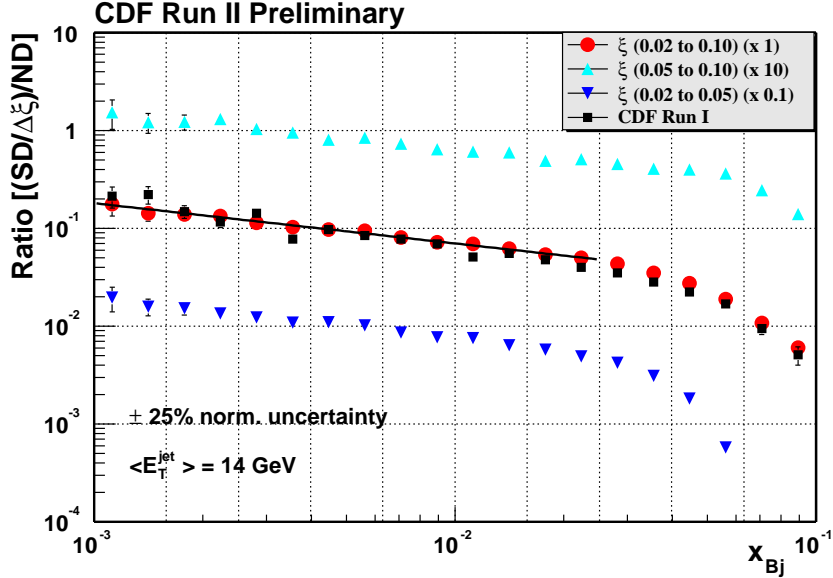,width=12cm}
\caption{Test of factorisation within CDF data alone.}
\end{center}
\label{fig14}
\end{figure}

\begin{figure}[t]
\begin{center}
\epsfig{file=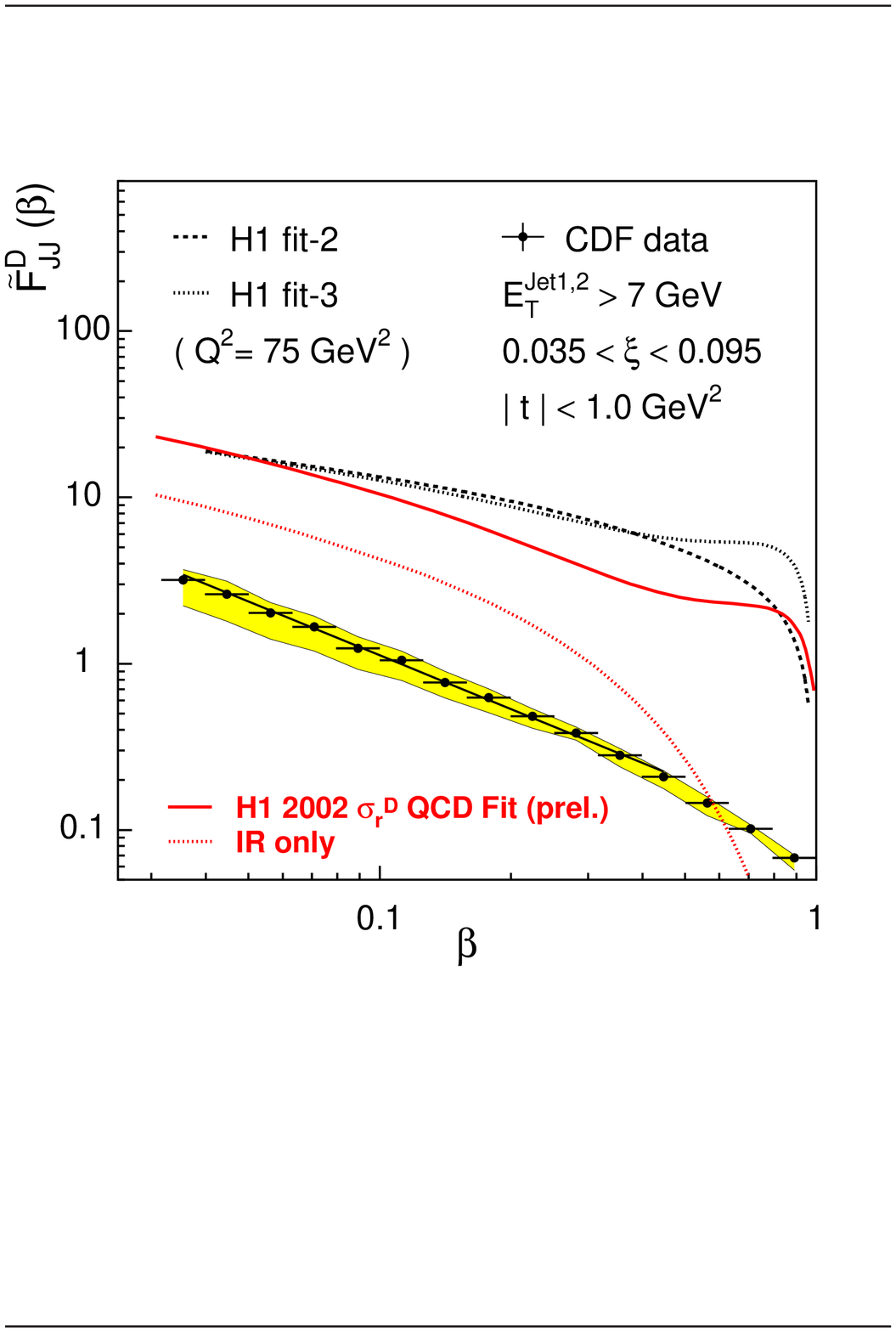,width=12cm,clip=true}
\vspace{12cm}
\caption{Comparison between the CDF measurement of diffractive structure
function (black points) with the expectation of the H1 QCD fits (red full line).}
\end{center}
\label{fig15}
\end{figure}

\begin{figure}[t]
\begin{center}
\epsfig{file=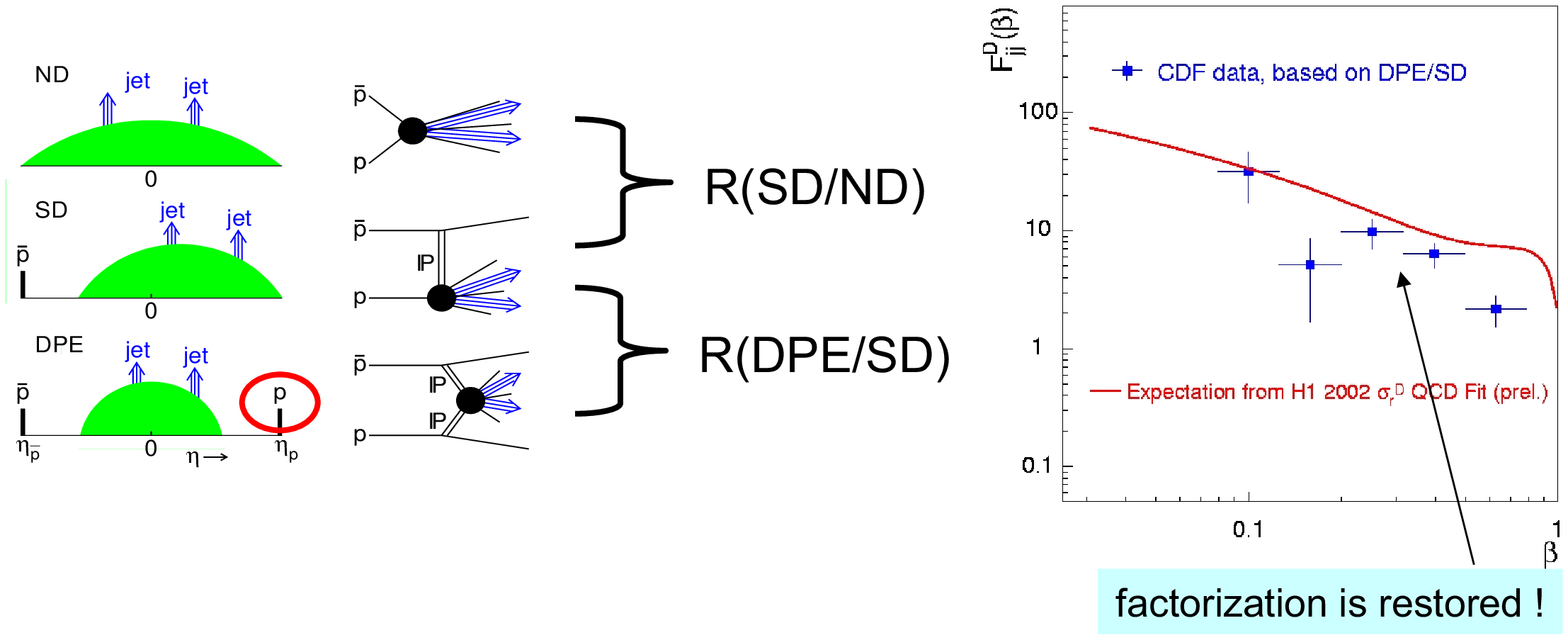,width=6cm,angle=270}
\caption{Restoration of factorisation for the ratio of double pomeron exchange
to single diffractive events (CDF Coll.).}
\end{center}
\label{fig16}
\end{figure}

\subsection{Survival probability studies in H1}
We mentioned in the previous section that the concept of survival probablity 
is related to soft gluon emission. This process can also be studied at HERA
using resolved photoproduction where events are sensitive to the hadronic
structure of the photon (see Fig. 17, right plot). The resolved process
is different from the direct one where the photon couples directly to the
pomeron (see Fig. 17, left plot).
In that case, we get an hadron hadron process like at
the Tevatron since we are sensitive to the hadronic contents of the photon.
In Fig. 18, we display the ratio between data
and NLO predictions for DIS (red triangles) and photoproduction data (black
points). We notice that we see a different of about a factor 2 between these two
data sets which might be an indication of survival probability effects. However,
no difference is observed between resolved or direct photoproduction 
where factorisation is expected to hold.

\begin{figure}[t]
\begin{center}
\begin{tabular}{cc}
\hspace{-1.5cm}
\epsfig{figure=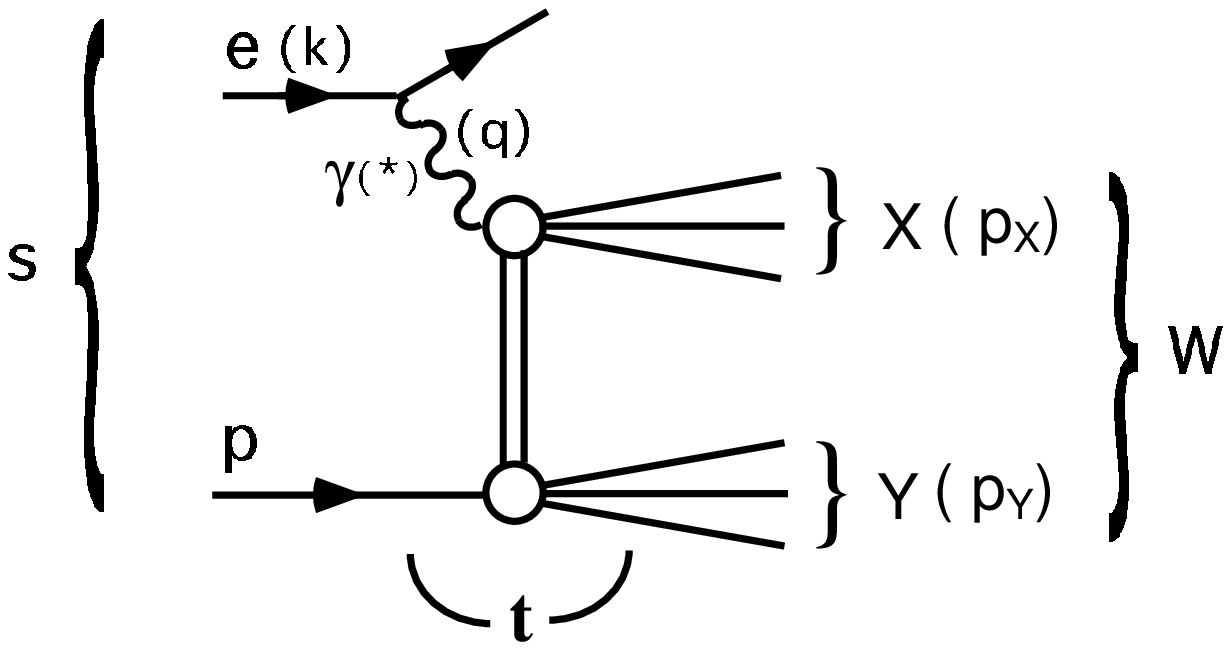,height=1.8in} &
\hspace{-0.5cm}
\epsfig{figure=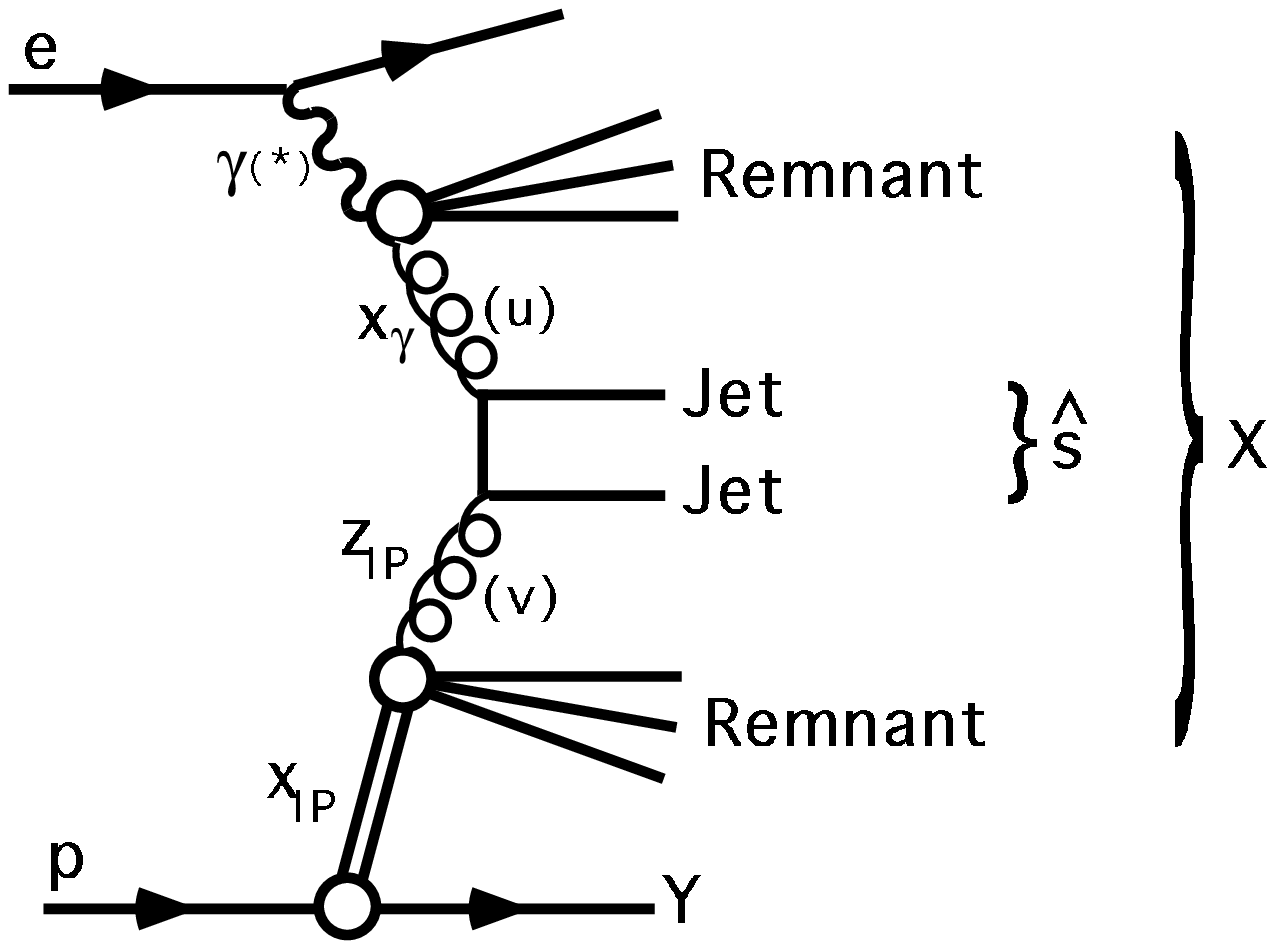,height=1.8in} \\
\end{tabular}
\caption{Scheme of direct (left) or resolved (right) photoproduction events.}
\end{center}
\label{fig15b}
\end{figure}

\begin{figure}[t]
\begin{center}
\epsfig{file=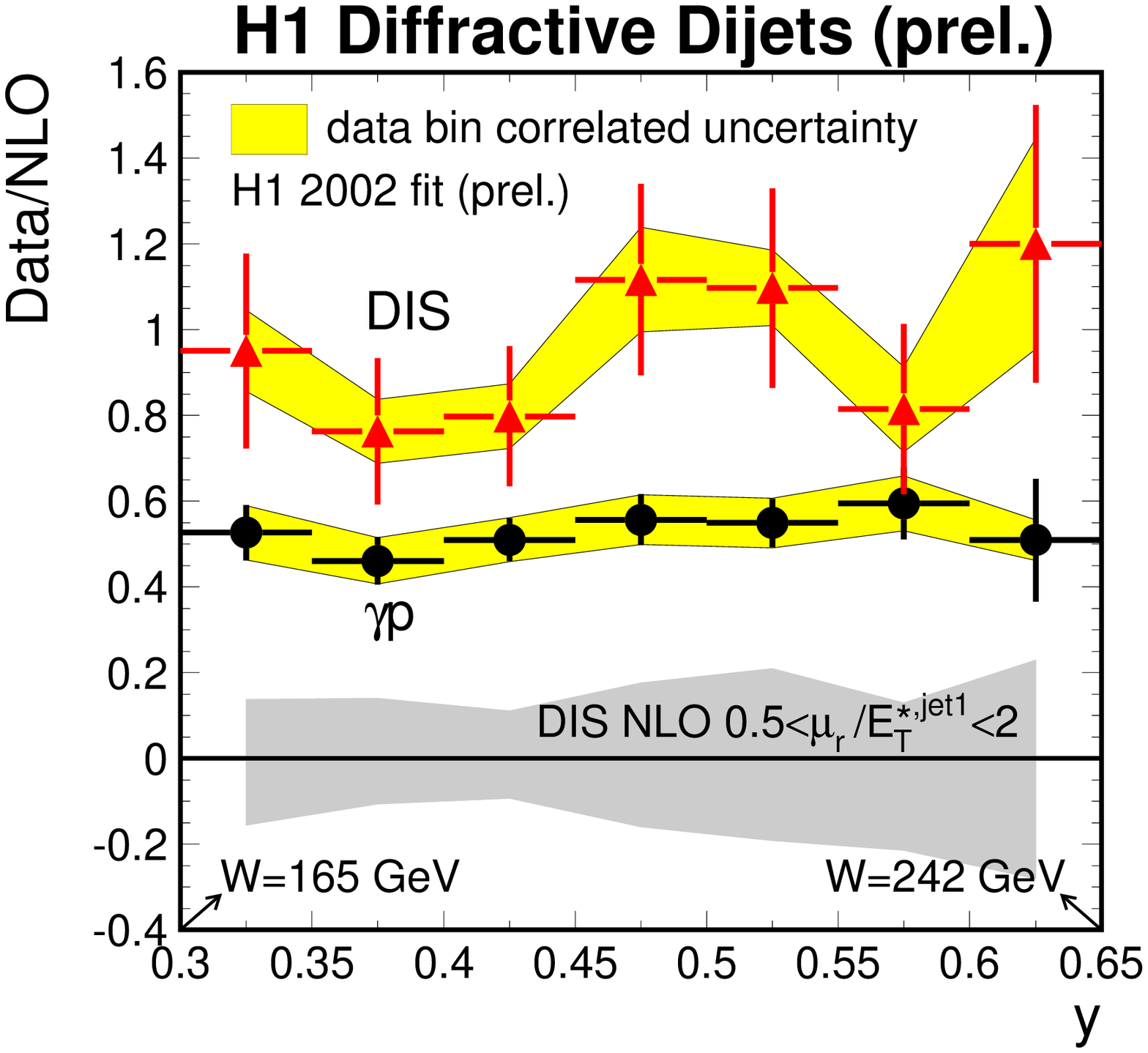,width=10cm}
\caption{Test of factorisation using photoproduction or DIS data in H1.}
\end{center}
\label{fig15c}
\end{figure}

\subsection{Possibility of survival probablity measurements at D\O\ }
A new measurement to be performed at the Tevatron,
in the D\O\ experiment has been proposed \cite{alexander}, which can be
decisive to test directly the concept of survival probability at the Tevatron, 
by looking at the azimuthal 
distributions of the outgoing proton and antiproton with respect to the 
beam direction. 

In Fig. 19, we display the survival probability for three different
values of $t$ as a function of the difference in azimuthal angle between the
scattered $p$ and $\bar{p}$. The upper black curve represents the case where the
$t$ of the $p$ and $\bar{p}$ are similar and close to 0. In that case, only a
weak dependence on $\Delta \Phi$ is observed. The conclusion is different for
asymmetric cases or cases when $t$ is different from 0: Fig. 19 
also shows the result in full red line for the asymmetric case ($t_1=0.2$,
$t_2=0.7$ GeV$^2$), and in full and dashed blue lines for $t_1=t_2=0.7$ GeV$^2$
for two different models of survival probabilities. We notice that we get a very
strong $ \Delta \Phi$ dependence of more than one order of magnitude. 

The $\Phi$ dependence can
be tested directly using the roman pot detectors at D\O\ (dipole
and quadrupole detectors) and their possibility
to measure the azimuthal angles of the $p$ and $\bar{p}$. For this purpose, 
we define the following configurations for dipole-quadrupole tags: 
same side (corresponding to $\Delta \Phi < 45$ degrees), 
opposite side (corresponding to $\Delta \Phi > 135$ degrees),
and middle side (corresponding to $45 < \Delta \Phi < 135$ degrees).
In Table 1, we give the 
ratios $middle/(2 \times same)$ and $opposite/same$ 
(note that we divide $middle$ by 2 to get the same domain size in
$\Phi$) for the different models. In order to obtain these predictions, we used
the full acceptance in $t$ and $\xi$ of the FPD detector.
Moreover the ratios for two  different tagging configurations,
namely for $\bar{p}$ tagged
in dipole detectors, and $p$ in quadrupoles, or for both $p$ and $\bar{p}$ 
tagged in quadrupole detectors \cite{alexander} were computed. 

The results are also compared
to expectations using another kind of model to describe diffractive events,
namely soft colour interaction \cite{sci}. This model assumes that diffraction
is not due to a colourless exchange at the hard vertex (called pomeron) but
rather to string rearrangement in the final state during hadronisation. In this
kind of model, there is a probability (to be determined by the experiment) that
there is no string connection, and so no colour exchange, between the partons
in the proton and the scattered quark produced during the hard interaction.
Since this model does not imply the existence of pomeron, there is no need of a
concept like survival probability, and no dependence on $\Delta \Phi$ of
diffractive cross sections. The proposed measurement would allow to distinguish
between these two dramatically different models of diffraction.

\begin{figure}[t]
\begin{center}
\epsfig{file=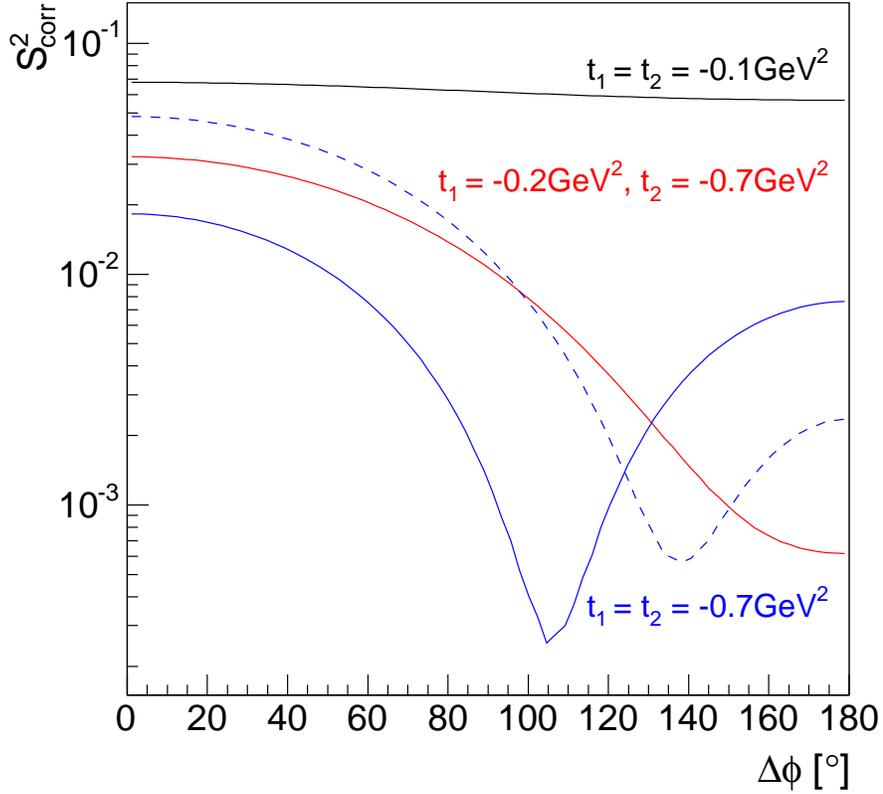,width=12cm}
\caption{$\Delta \Phi$ dependence of the survival probability for two different
models of survival probability where $\Delta \Phi$ is the difference in
azimuthal angle between the scattered $p$ and $\bar{p}$ in the final state, and
for three different values of $t$ (see text).}
\end{center}
\label{fig16}
\end{figure}

\begin{table}
\begin{center}
\begin{tabular}{|c|c||c|c|} \hline
 Config. & model & midd./ & opp./ \\ 
         &       &  same  & same \\
\hline\hline
Quad.  & SCI & 1.3 & 1.1 \\
 $+$ Dipole              & Pom. 1 & 0.36 & 0.18 \\
               & Pom. 2 & 0.47 & 0.20 \\ \hline
Quad.  & SCI & 1.4 & 1.2 \\
$+$ Quad.               & Pom. 1 & 0.14 & 0.31 \\
               & Pom. 2 & 0.20 &  0.049     
\\
\hline
\end{tabular}
\end{center}
\caption{Predictions for a proposed measurement of diffractive 
cross section ratios in different regions of $\Delta \Phi$ at the Tevatron
(see text for the definition of middle, same and opposite).
The first (resp. second) measurement involves the quadrupole and dipole
detectors (resp. quadrupole detectors only) leading to asymmetric (resp.
symmetric) cuts on $t$.
}
\end{table}

\section{Diffractive exclusive event production}

\subsection{Interest of exclusive events}
A schematic view of non diffractive, inclusive double pomeron exchange,
exclusive diffractive events at the Tevatron or the LHC is displayed in Fig.
20. The upper left plot shows the ``standard" non diffractive events
where the Higgs boson, the dijet or diphotons are produced directly by a
coupling to the proton and shows proton remnants. The bottom plot displays 
the standard diffractive double
pomeron exchange where the protons remain intact after interaction and the total
available energy is used to produce the heavy object (Higgs boson, dijets,
diphotons...) and the pomeron remnants. We have so far only discussed
this kind of events and their diffractive production using the
parton densities measured at HERA. There may be a third class of processes
displayed in the upper right figure, namely the exclusive diffractive
production. In this kind of events, the full energy is used to produce the heavy
object (Higgs boson, dijets, diphotons...) and no energy is lost in pomeron
remnants. There is an important kinematical consequence: the mass of the
produced object can be computed using roman pot detectors and tagged protons:

\begin{eqnarray}
M = \sqrt{\xi_1 \xi_2 S}.
\end{eqnarray} 
We see immediately the advantage of those processes: we can benefit from the
good roman pot resolution on $\xi$ to get a good resolution on mass. It is then
possible to measure the mass and the kinematical properties of the produced
object and use this information to increase the signal over background ratio by reducing the
mass window of measurement. It is thus important to know if this kind of events
exist or not.

\begin{figure}[t]
\begin{center}
\epsfig{file=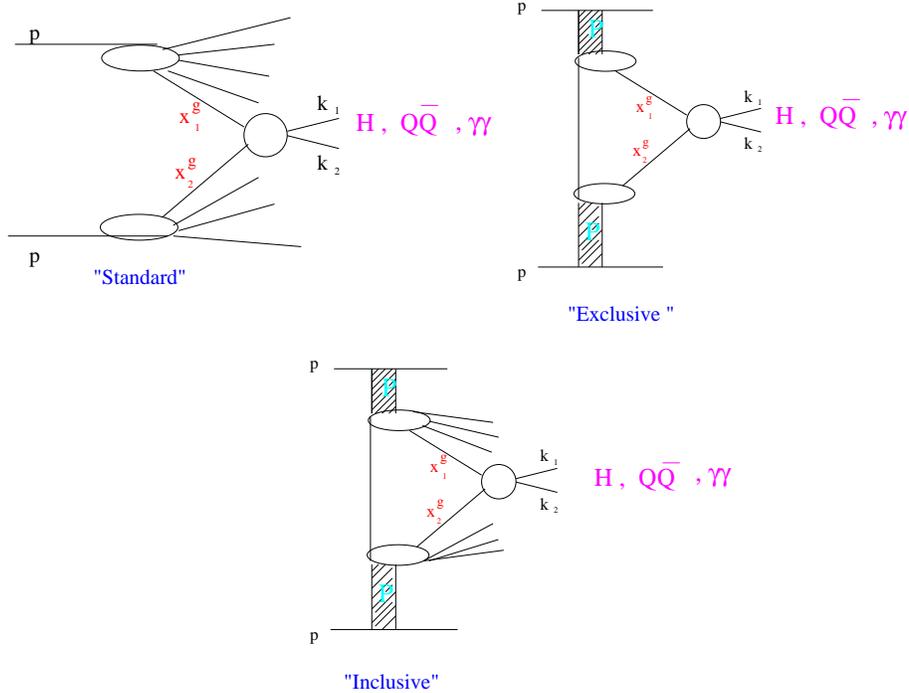,width=12cm}
\caption{Scheme of non diffractive, inclusive double pomeron exchange,
exclusive diffractive events at the Tevatron or the LHC.}
\end{center}
\label{fig17}
\end{figure}

\subsection{Search for exclusive events at the Tevatron}

The CDF collaboration measured the so-called dijet mass fraction
in dijet events - the ratio of the mass carried by the two jets divided by the
total diffractive mass - when the antiproton is tagged in the roman pot
detectors and when there is a rapidity gap on the proton side to ensure that the
event corresponds to a double pomeron exchange. The results are shown in Fig. 
21 and are compared  with the POMWIG \cite{pomwig} expectation using the gluon and
quark densities measured by the H1 collaboration in dashed line \cite{cdfdiff}. We 
see a clear deficit of
events towards high values of the dijet mass fraction, where exclusive events
are supposed to occur (for exclusive events, the dijet mass fraction is 1 by
definition at generator level and can be smeared out towards lower values taking
into account the detector resolutions). Fig. 21 shows also the
comparison between data and the predictions from the POMWIG and DPEMC
generators,
DPEMC being used to generate exclusive events \cite{ushiggs}. There is
a good agreement between data and MC. However, this does not prove the existence of
exclusive events since the POMWIG prediction shows large uncertainties (the
gluon in the pomeron used in POMWIG is not the latest one obtained by the H1
collaboration \cite{h1f2d, us} and the 
uncertainty at high $\beta$ is quite large as we discussed in a previous
section). The results (and the conclusions) might change using the newest gluon
density and will be of particular interest.
In addition, it is not obvious one
can use the gluon density measured at HERA at the Tevatron since factorisation
does not hold, or in other words, this assumes that the survival probability is a
constant, not depending on the kinematics of the interaction. 

A direct precise
measurement of the gluon density in the pomeron through the measurement of the
diffractive dijet cross section at the Tevatron and the LHC will be necessary
if one wants to prove the existence of exclusive events in the dijet channel.
However, this measurement is not easy and requires a full QCD analysis. We
expect that exclusive events would appear as a bump in the gluon distribution at
high $\beta$, which will be difficult to interprete. To show that this bump is
not due to tail of the inclusive distribution but real exclusive events, it
would be necessary to show that those tails are not compatible with a standard
DGLAP evolution of the gluon density in the pomeron as a function of jet 
transverse momentum. However, it does not seem to be easy to distinguish those
effects from higher twist ones. It is thus important to look for different
methods to show the existence of exclusive events.

\begin{figure}[t]
\begin{center}
\epsfig{file=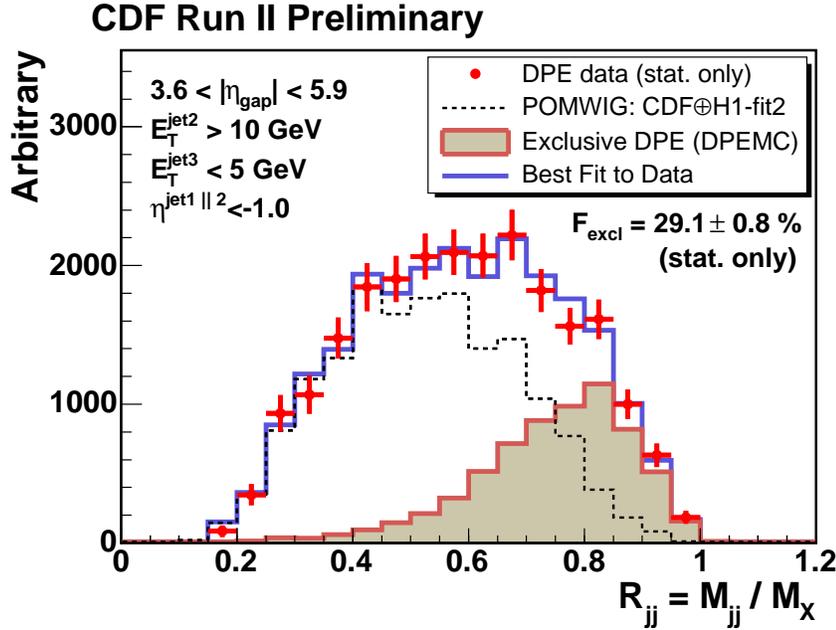,width=12cm}
\caption{Search for exclusive diffractive events at CDF.}
\end{center}
\label{fig17}
\end{figure}

The CDF collaboration also looked for the exclusive production of dilepton and
diphoton. Contrary to diphotons, dileptons cannot be produced exclusively via pomeron exchanges since
$g g \rightarrow \gamma \gamma$ is possible, but $g g \rightarrow l^+ l^-$ 
directly is impossible. However, dileptons can be produced via QED processes, and
the cross section is perfectly known. The CDF measurement is $\sigma = 1.6
^{+0.5}_{-0.3} (stat) \pm 0.3 (syst)$ pb which is found to be in good agreement
with QED predictions and shows that the acceptance, efficiencies of the detector
are well understood. Three exclusive diphoton events have been observed by the CDF
collaboration leading to a cross section of
$\sigma = 0.14
^{+0.14}_{-0.04} (stat) \pm 0.03 (syst)$ pb compatible with the expectations
for exclusive diphoton production at the Tevatron.

Other searches like $\chi_C$ production and the ratio of diffractive $b$ jets to
the non diffractive ones as a function of the dijet mass fraction show further
indications that exclusive events might exist but there is no definite proof
until now.

\subsection{Search for exclusive events at the LHC}
The search for exclusive events at the LHC can be performed in the same channels
as the ones used at the Tevatron. In addition, some other possibilities
benefitting from the high luminosity of the LHC appear. One of the cleanest way
to show the existence of exclusive events would be to measure the dilepton and
diphoton cross section ratios as a function of the dilepton/diphoton mass. If
exclusive events exist, this distribution should show a bump towards high values
of the dilepton/diphoton mass since it is possible to produce exclusively
diphotons but not dileptons at leading order as we mentionned in the previous
paragraph. 

The search for exclusive events at the LHC will also require a precise analysis
and measurement of inclusive diffractive cross sections and in particular the
tails at high $\beta$ since it is a direct
background to exclusive event production.

\section{Diffraction at the LHC}
In this section, we will describe briefly some projects concerning diffraction
at the LHC. We will put slightly more emphasis on the diffractive production
of heavy objects such as Higgs bosons, top  or stop pairs, $WW$ events...

\subsection{Diffractive event selection at the LHC}
The LHC with a center-of-mass energy of 14 TeV will allow us to access a completely
new kinematical domain in diffraction. So far, two experiments, namely ATLAS and
CMS-TOTEM have shown interests in diffractive measurements.
The diffractive event selection at the LHC will be the same as at the Tevatron.
However, the rapidity gap selection will no longer be possible at high
luminosity since up to 25 interactions per bunch crossing  are expected to occur
and soft pile-up events will kill the gaps produced by the hard interaction.
Proton tagging will thus be the only possibility to detect diffractive events at
high luminosity.

\subsection{Measurements at the LHC using a high $\beta^*$ lattice}
Measurements of total cross section and luminosity are foreseen in the ATLAS
\cite{atlaslumi} and TOTEM  \cite{totem} experiments, and roman pots are
installed at 147 and 220 m in TOTEM and 240 m in ATLAS. These measurements will
require a special injection lattice of the LHC at low luminosity since they require the
roman pot detectors to be moved very close to the beam. As an example, the
measurement of the total cross section to be performed by TOTEM \cite{totem} is shown in Fig.
22. We notice that there is a large uncertainty on prediction of the total cross
section at the LHC energy in particular due to the discrepancy between the two
Tevatron measurements, and this measurement of TOTEM will be of special
interest.

\begin{figure}[t]
\begin{center}
\epsfig{file=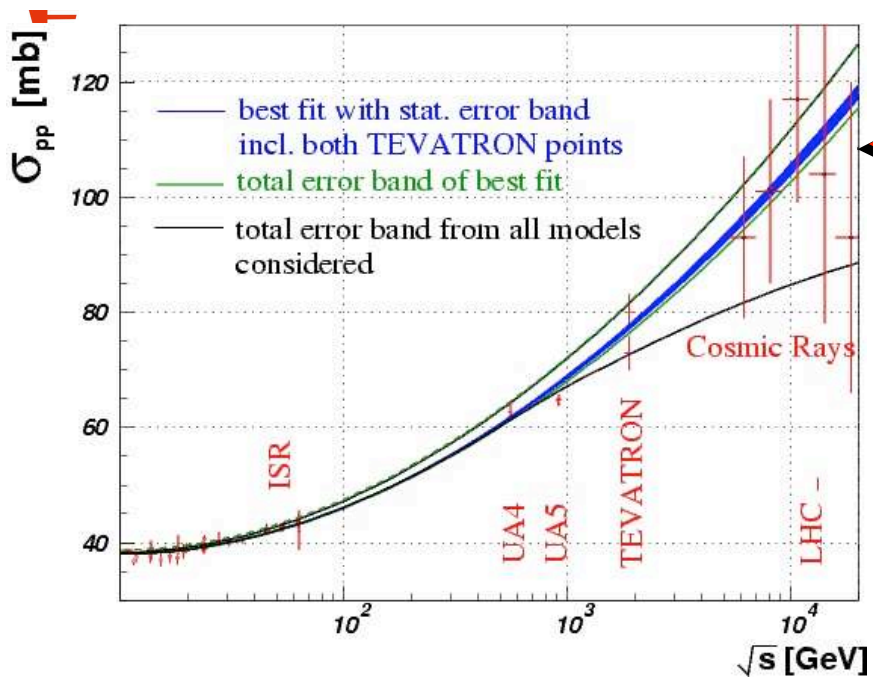,width=10cm,angle=270}
\caption{Measurement of the total cross section.}
\end{center}
\label{fig18}
\end{figure}

\subsection{Hard inclusive diffraction at the LHC}
In this section, we would like to discuss how we can measure the gluon density
in the pomeron, especially at high $\beta$ since the gluon in this kinematical 
domain shows large uncertainties and this is where the exclusive contributions
should show up if they exist. To take into account the high-$\beta$
uncertainties of the gluon distribution, we chose to multiply the gluon density
in the pomeron measured at HERA by a factor $(1-\beta)^{\nu}$  where $\nu$
varies between -1.0 and 1.0. If $\nu$ is negative, we enhance the gluon density
at high $\beta$ by definition, especially at low $Q^2$.

A possible measurement at the LHC is described in Fig. 23. The dijet
mass fraction is shown in dijet diffractive production  for different jet 
transverse momenta  ($P_T>$ 100 (upper left), 200 (upper right),
300 (lower left) and 400 GeV (lower right)), and for the different values if
$\nu$. We notice that the variation of this distribution as a function of jet
$p_T$ can assess directly the high $\beta$ behaviour of the gluon density.
In the same kind of ideas, it is also possible to use $t \bar{t}$ event
production to test the high-$\beta$ gluon. Of course, this kind of measurement
will not replace a direct QCD analysis of the diffractive dijet cross section
measurement.

Other measurements already mentionned such as the diphoton, dilepton cross
section ratio as a function of the dijet mass, the $b$ jet, $\chi_C$, $W$ and
$Z$  cross section measurements will be also quite important at the LHC.

\begin{figure}[t]
\begin{center}
\begin{tabular}{cc}
\hspace{-1.cm}
\epsfig{figure=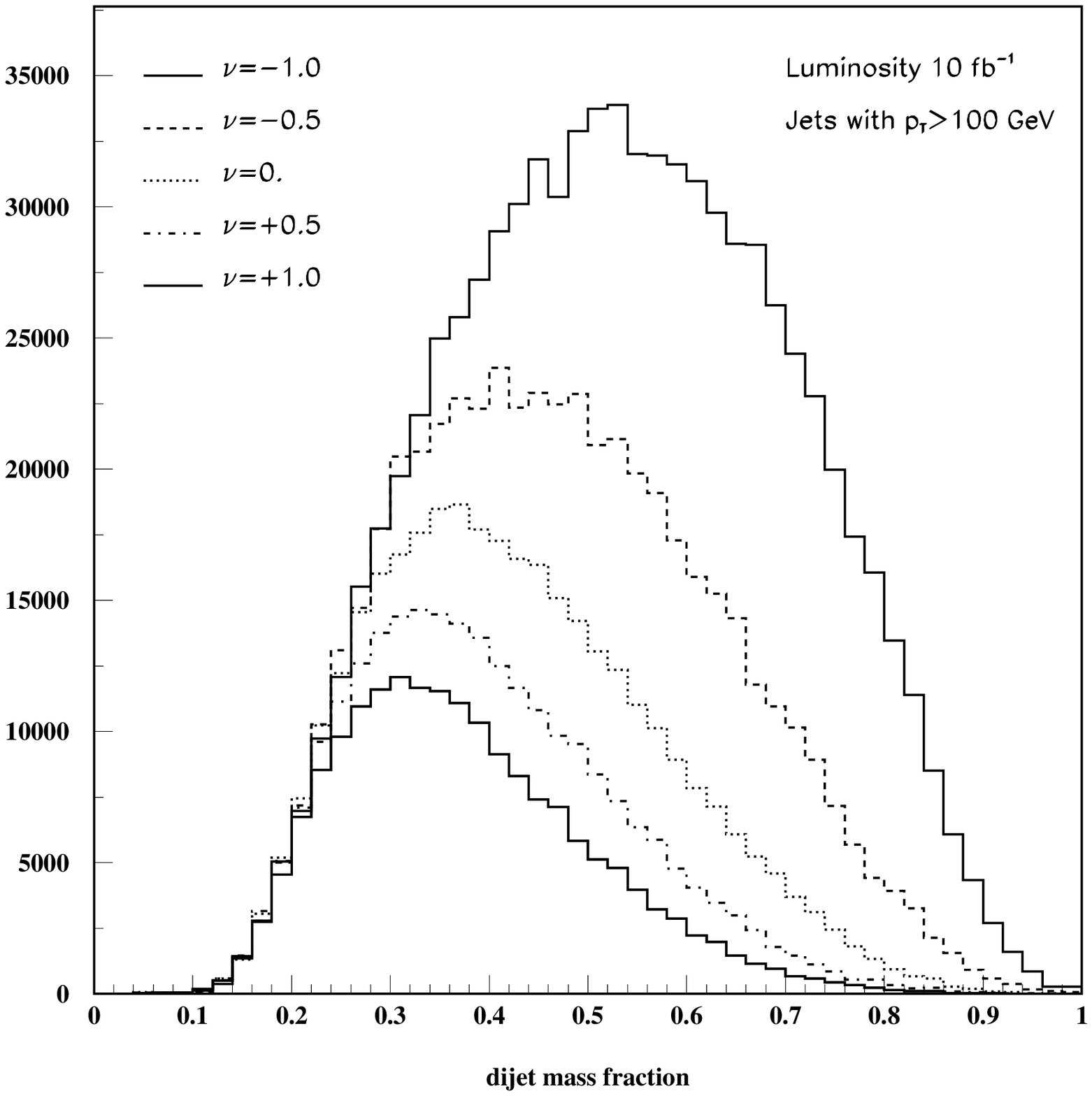,height=2.5in} &
\hspace{-0.5cm}
\epsfig{figure=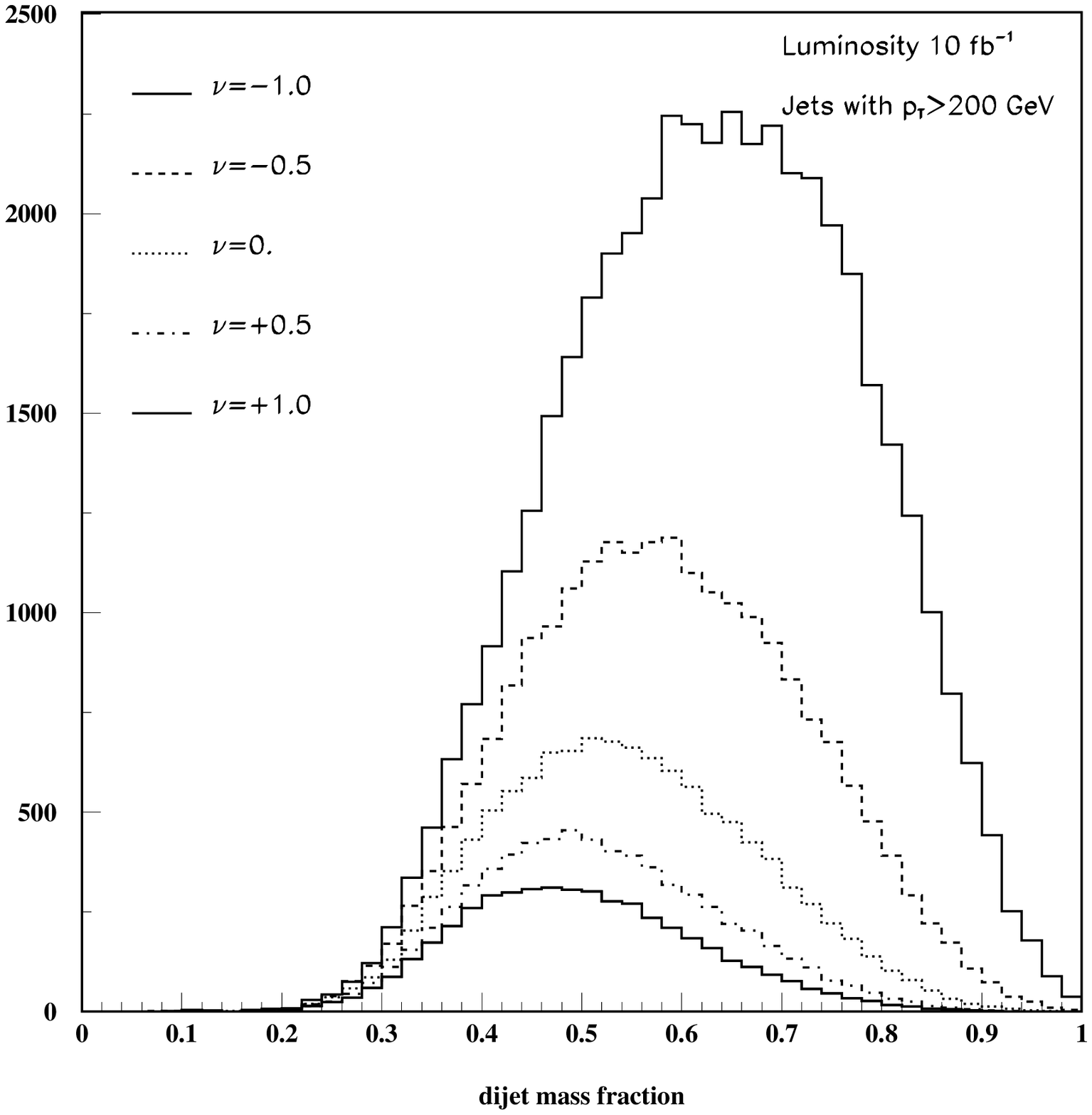,height=2.5in} \\
\hspace{-1.cm}
\epsfig{figure=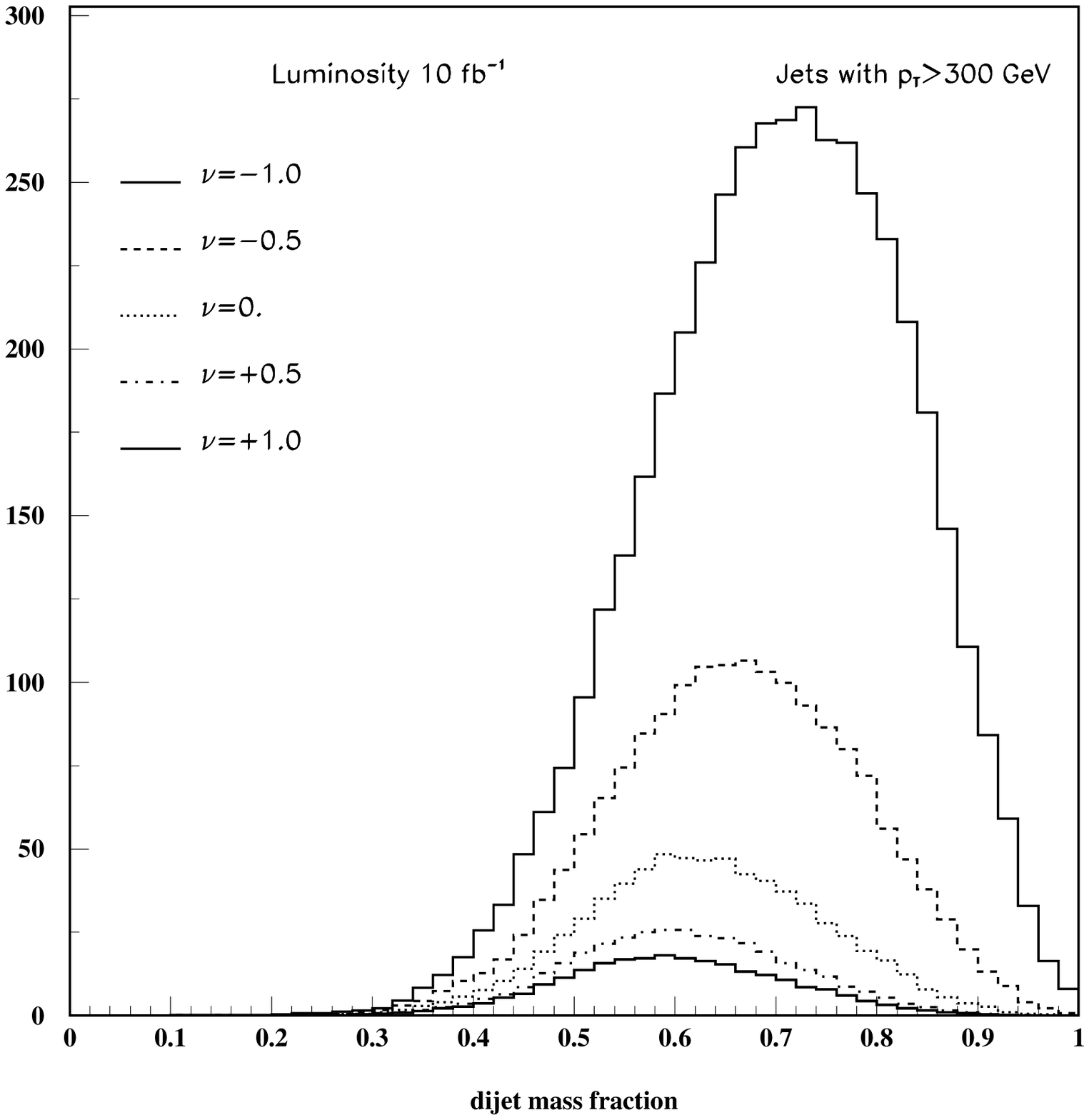,height=2.5in} &
\hspace{-0.5cm}
\epsfig{figure=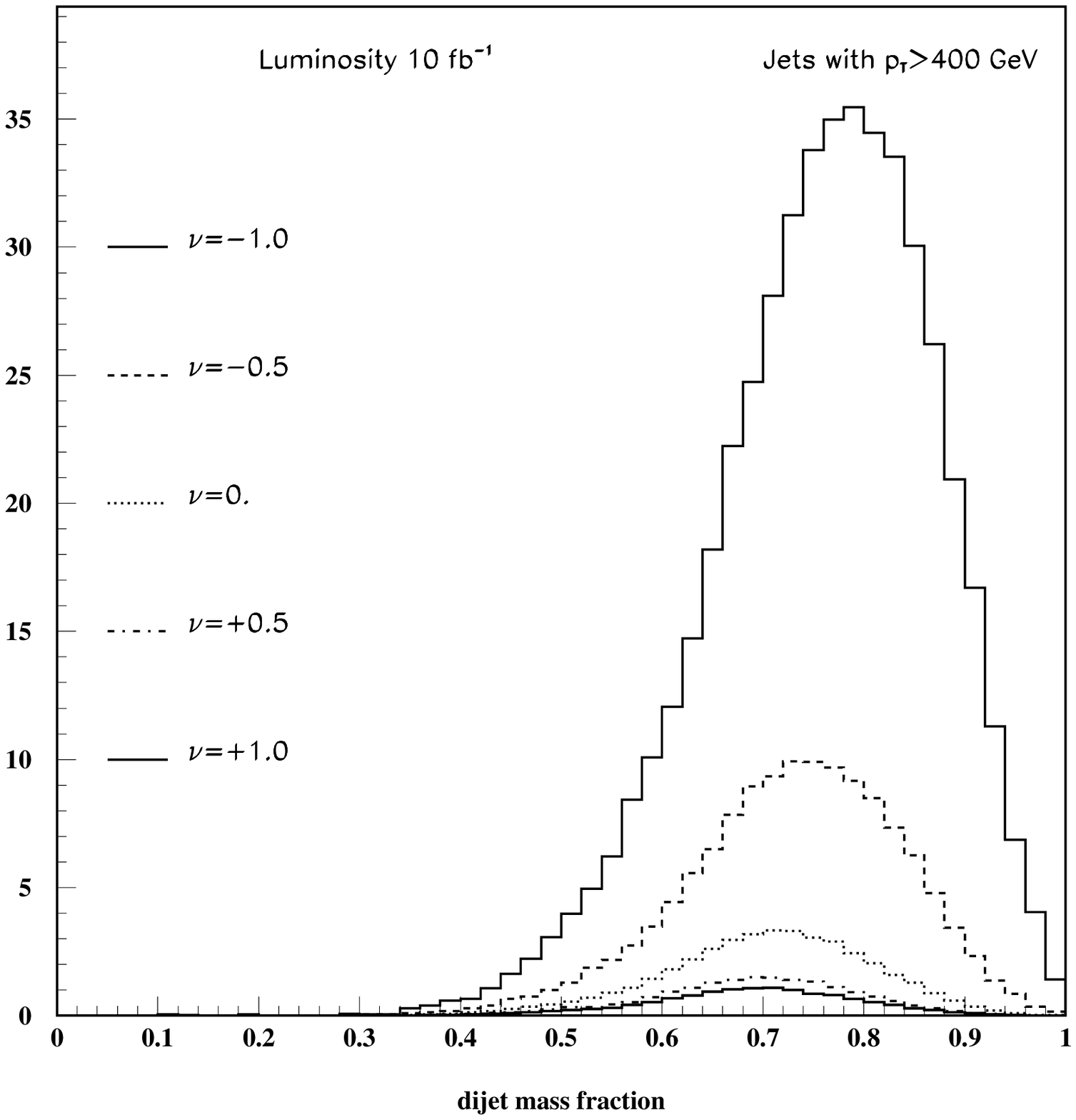,height=2.5in} \\
\end{tabular}
\caption{Dijet mass fraction for jet $P_T>$ 100 (upper left), 200 (upper right),
300 (lower left) and 400 GeV (lower right) for different gluon assumptions at
high $\beta$ (the gluon is multiplied by $(1-\beta)^{\nu}$).}
\end{center}
\label{fig19}
\end{figure}

\subsection{Exclusive Higgs production at the LHC}
As we already mentionned in one of the previous sections, one special interest of
diffractive events at the LHC is related to the existence of exclusive events.
So far, two projects are being discussed at the LHC: the installation of roman
pot detectors at 220 m in ATLAS \cite{atlas}, and at 420 m for the ATLAS and CMS
collaborations \cite{fp420}. 

The results discussed in this section rely on the DPEMC Monte Carlo to produce 
Higgs bosons exclusively \cite{ushiggs, dpemc} and a fast simulation of a
typical LHC detector (ATLAS or CMS). 
Results are given in Fig. 24 for a Higgs mass of 120 GeV, 
in terms of the signal to background 
ratio S/B, as a function of the Higgs boson mass resolution.
Let us notice that the background is mainly due the exclusive $b \bar{b}$
production. However the tail of the inclusive $b \bar{b}$ production can also be
a relevant contribution and this is related to the high $\beta$ gluon
density which is badly known at present. 
In order to obtain a S/B of 3 (resp. 1, 0.5), a mass resolution of about
0.3 GeV (resp. 1.2, 2.3 GeV) is needed. A mass resolution of the order of 1 GeV
seems to be technically feasible.

The diffractive SUSY Higgs boson production cross section is noticeably enhanced 
at high values of $\tan \beta$ and since we look for Higgs decaying into $b
\bar{b}$, it is possible to benefit directly from the enhancement of the cross
section contrary to the non diffractive case. A signal-over-background up to a
factor 50 can be reached for 100 fb$^{-1}$ for $\tan \beta \sim 50$
\cite{lavignac} (see Fig. 25).

\begin{table}
\begin{center}
\begin{tabular}{|c||c|c|c|c|c|} \hline
$M_{Higgs}$& cross & signal & backg. & S/B & $\sigma$  \\
 & section &  & & & \\
\hline\hline
120 & 3.9 & 27.1 & 28.5 & 0.95 & 5.1  \\
130 & 3.1 & 20.6 & 18.8 & 1.10 & 4.8  \\
140 & 2.0 & 12.6 & 11.7 & 1.08 & 3.7  \\ 
\hline
\end{tabular}
\caption{Exclusive Higgs production cross section for different Higgs masses,
number of signal and background events for 100 fb$^{-1}$, ratio, and number of
standard deviations ($\sigma$).}
\label{sb}
\end{center}
\end{table}

\begin{figure}[t]
\begin{center}
\epsfig{file=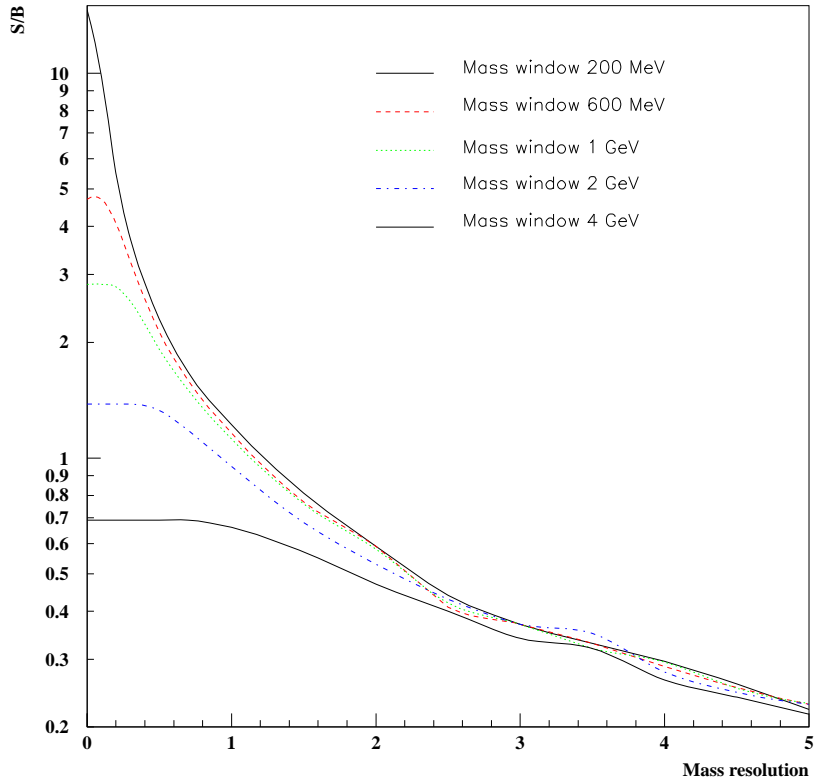,width=12cm}
\caption{Standard Model Higgs boson signal to background ratio as a function 
of the resolution on the missing mass, in GeV. This figure assumes a Higgs
boson mass of 120 GeV.}
\end{center}
\label{fig20}
\end{figure}

\begin{figure}[t]
\begin{center}
\epsfig{file=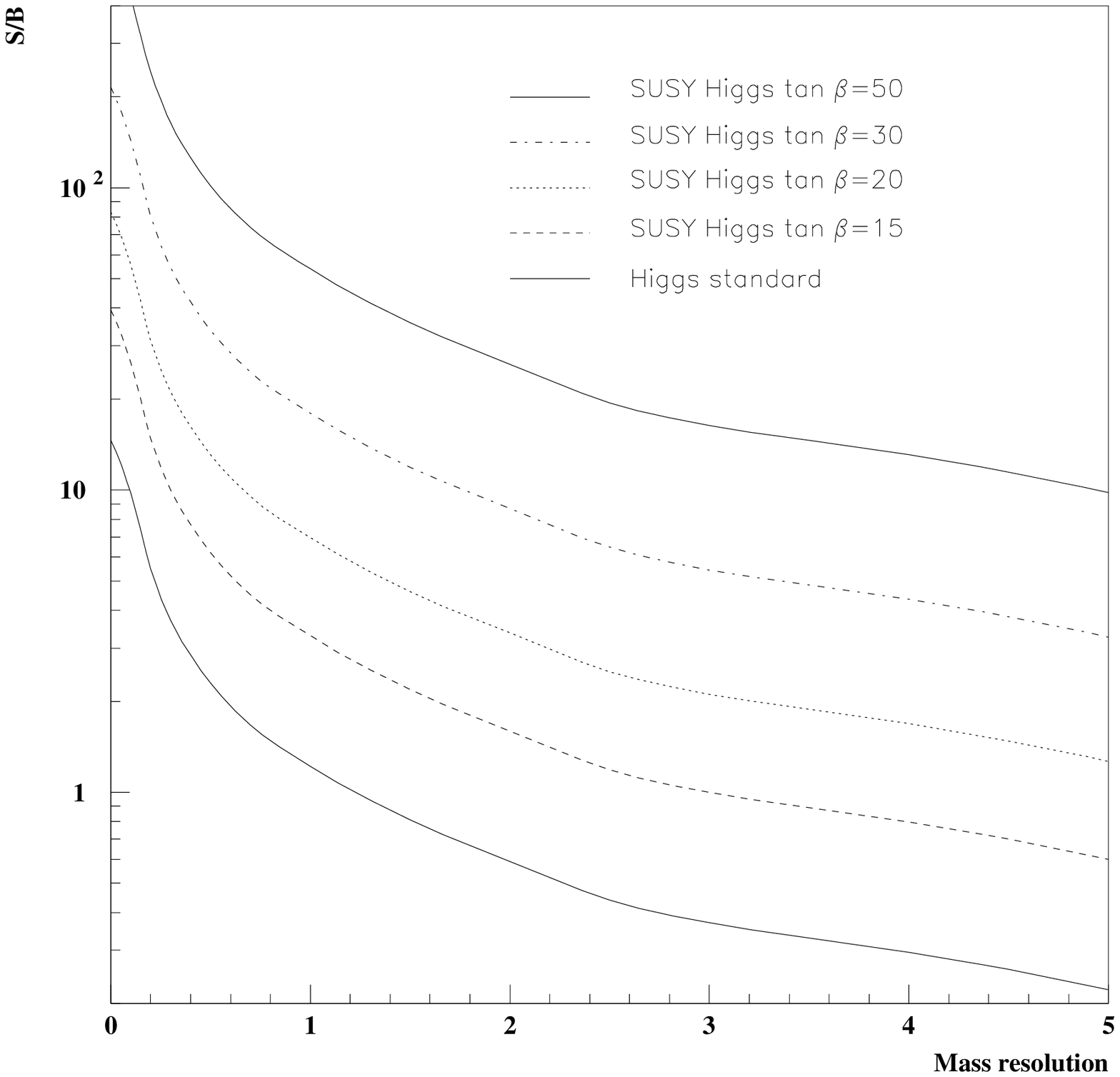,width=12cm}
\caption{SUSY Higgs boson signal to background ratio as a function 
of the resolution on
the missing mass, in GeV. This figure assumes a Higgs
boson mass of 120 GeV.}
\end{center}
\label{fig20b}
\end{figure}

\subsection{Exclusive top, stop and $W$ pair production at the LHC}
In the same way that Higgs bosons can be produced exclusively, it is possible 
to produce $W$, top and stops quark pairs.
$WW$ bosons are produced via QED processes which means that their cross section
is perfectly known. On the contrary, top and stop pair production are obtained via
double pomeron exchanges and the production cross section is still uncertain.

The method to reconstruct the mass of heavy objects
double diffractively produced at the LHC is
based on a fit to the turn-on point of the missing mass distribution at 
threshold \cite{jochen}. 

One proposed method (the ``histogram'' method) corresponds to the comparison of 
the mass distribution in data with some reference distributions following
 a Monte Carlo simulation of the detector with different input masses
corresponding to the data luminosity. As an example, we can produce 
a data sample for 100 fb$^{-1}$ with a top mass of 174 GeV, and a few 
MC samples corresponding to different top masses between 150 and 200 GeV. 
For each Monte Carlo sample, a $\chi^2$ value corresponding to the 
population difference in each bin between data and MC is computed. The mass point 
where
the $\chi^2$ is minimum corresponds to the mass of the produced object in data.
This method has the advantage of being easy but requires a good
simulation of the detector.

The other proposed method (the ``turn-on fit'' method) is less sensitive to the MC 
simulation of the
detectors. As mentioned earlier, the threshold scan is directly sensitive to
the mass of the diffractively produced object (in the $WW$ case for instance, it
is sensitive to twice the $WW$ mass). The idea is thus to fit the turn-on
point of the missing mass distribution which leads directly to the mass 
of the produced object, the $WW$ boson. Due to its robustness,
this method is considered as the ``default" one.

The precision of the $WW$ mass measurement (0.3 GeV for 300 fb$^{-1}$) is not 
competitive with other 
methods, but provides a very precise check of the calibration 
of the roman pot detectors. $WW$ events will also allow to assess directly the
sensitivity to the photon anomalous coupling since it would reveal itself by a
modification of the well-known QED $WW$ production cross section. We can notice
that the $WW$ production cross section is proportional to the fourth power of
the $\gamma W$ coupling which ensures a very good sensitivity of that process
\cite{olda}.
The precision of
the top mass measurement is however competitive, with an expected precision 
better than 1 GeV at high luminosity provided that the cross section
is high enough. 
The other application is to use the so-called ``threshold-scan method"
to measure the stop mass \cite{lavignac}. After taking into account the stop width, we obtain a resolution
on the stop mass of 0.4, 0.7 and 4.3 GeV for a stop mass of 174.3, 210 and 393
GeV for a luminosity (divided by the signal efficiency) of 100 fb$^{-1}$. 

The caveat is  of course that the production via diffractive 
exclusive processes is model dependent, and definitely needs
the Tevatron and LHC data to test the models. It will allow us to determine more precisely 
the production cross section by testing and measuring at the Tevatron the jet 
and photon production for high masses and high dijet or diphoton mass fraction.

\section{Conclusion}
In these lectures, we presented and discussed the most recent results on
inclusive diffraction from the HERA and Tevatron experiments and gave the
prospects for the future at the LHC. Of special interest is the exclusive
production of Higgs boson and heavy objects ($W$, top, stop pairs) which will
require a better understanding of diffractive events and the link between $ep$
and hadronic colliders, and precise measurements and analyses of inclusive
diffraction at the LHC in particular to constrain further the gluon density in
the pomeron.

\section*{Acknowledgments}
I thank Robi Peschanski and Old\v{r}ich Kepka for a careful reading of the
manuscript.


\end{document}